\documentclass[aps,prd,twocolumn,showpacs,superscriptaddress,groupedaddress]{revtex4-1}  
\usepackage{graphicx}  
\usepackage{dcolumn}   
\usepackage{bm}        
\usepackage{amssymb}   
\usepackage{amsmath}   

\usepackage{color} 

\usepackage{multirow}
\usepackage{enumerate}
\usepackage{comment}
\usepackage{tabularx}
\usepackage{booktabs}

\newcommand{\irm}{{\mathrm i}}

\begin{document}

\title{Impact of infrasound atmospheric noise on gravity detectors used for astrophysical and geophysical applications}

\author{Donatella Fiorucci$^{1}$} 
\author{Jan Harms$^{2,3}$}
\author{Matteo Barsuglia$^{1}$} 
\author{Irene Fiori$^{4}$} 
\author{Federico Paoletti$^{4,5}$} 

\address{$^{1}$APC, AstroParticule et Cosmologie, Universit\'e Paris Diderot, CNRS/IN2P3, CEA/Irfu, Observatoire de Paris, Sorbonne Paris Cit\'e, 10, rue Alice Domon et L\'eonie Duquet, 75205 Paris Cedex 13, France}
\address{$^{2}$Gran Sasso Science Institute (GSSI), I-67100 L'Aquila, Italy}
\address{$^{3}$INFN, Laboratori Nazionali del Gran Sasso, I-67100 Assergi, Italy}
\address{$^{4}$European Gravitational Observatory (EGO), I-56021 Cascina, Pisa, Italy}
\address{$^{5}$INFN, Sezione di Pisa, I-56127 Pisa, Italy}
\date{\today}

\begin{abstract}

Density changes in the atmosphere produce a fluctuating gravity field that affect gravity strainmeters or gravity gradiometers used for the detection of gravitational-waves and for geophysical applications. This work addresses the impact of the atmospheric local gravity noise on such detectors, extending previous analyses. In particular we present the effect introduced by the building housing the detectors, and we analyze local gravity-noise suppression by constructing the detector underground. We present also new sound spectra and correlations measurements. The results obtained are important for the design of future gravitational-wave detectors and gravity gradiometers used to detect prompt gravity perturbations from earthquakes.  

\end{abstract}

\maketitle
\section{Introduction}

Ground acceleration noise enforces a fundamental sensitivity limitation to gravimeters. This limit can be overcome by realizing a differential readout between two test masses. The corresponding measuring device is called gravity gradiometer or gravity strainmeter. 

These devices are interesting for at least two applications. The first is the detection of gravitational-waves, ripples in space-time produced by acceleration of asymmetric mass distributions. Gravitational waves produced by binary black-hole and binary neutron-star mergers have been detected~\cite{AbEA2016a,AbEA2016d,AbEA2017b,AbEA2017c,AbEA2017d} by the LIGO-Virgo~\cite{LSC2015, AcEA2015} detectors, a network of modified Michelson interferometers with several kilometer-long arms and suspended test masses. These large-scale detectors are sensitive to gravity fluctuations between about 10\,Hz and a few 1000\,Hz.

The second application of gravity strainmeters is the measurement of geophysical signals such as the prompt gravity signal due to an earthquake as was recently proposed \cite{HaEA2015}. The first evidence of this signal was found in 2016, analysing the data recorded during the Japanese Tohoku-Oki 2011 earthquake~\cite{MoEA2016}. Since changes of the gravity field propagate at the speed of light, much faster than the seismic waves, a network of gravity gradiometers has been proposed for a faster determination of the earthquake magnitude and to improve the current earthquake early-warning systems~\cite{MoEA2016, HaEA2015}. Geophysical applications require gravity observations well below 10\,Hz ($\sim$ 0.01\,Hz - $\sim$ 1\,Hz). 

One of the predicted sensitivity limitations of gravity strainmeters is given by gravity fluctuations due to density perturbations in the vicinity of the sensor. The corresponding contribution to the instrumental noise is also called \textit{Newtonian Noise} (NN) in the gravitational-wave community~\cite{Sau1984}, to distinguish it from the gravitational waves, which are a purely relativistic effect. There are two main components of NN: the seismic NN, given by the density changes in the ground produced by seismic waves, and the atmospheric NN, given by the air density changes produced by pressure and temperature fluctuations. This paper addresses the atmospheric density changes. 

This topic has been treated mainly in~\cite{Har2015, Cre2008}, and only for gravitational-wave detectors operating at frequencies higher than a few Hz (as for instance Virgo and LIGO). Here we will extend the analysis of a particular type of atmospheric NN, the infrasound noise, to the detectors operating at frequencies below 1\,Hz (as, for instance, torsion bar antennas), used either for gravitational-wave detection or for geophysical applications. Moreover, we will complete the estimation of noise spectra in the following ways. First, we study the effect of the building housing the detector. Second, we analyze the possibility of attenuating this noise by placing the detector underground. Third, we present a new measurement of sound spectra necessary for the estimation of infrasound NN. 

The text is organized as follows: In section~\ref{sec:strainmeters}, we briefly describe the currently existing gravity strainmeters for gravity observations above 0.1\,Hz. In section \ref{sec:sources}, we introduce and describe the different sources of atmospheric NN as presented in previous works. In section \ref{sec:inm}, we give the details of the mathematical framework we used to model the infrasound noise for the different configurations we considered. In section \ref{sec:sim}, we describe the numerical simulation used to compute the infrasound NN levels. In section \ref{sec:pspectra}, we report on the sound measurements. In section \ref{sec:results},  we summarize the main results of this study, and finally in section \ref{sec:conclusions}, we give the conclusions and the outlook of this work.

\section{Gravity strainmeters for astrophysical and geophysical applications}
\label{sec:strainmeters}

\begin{figure}
  \centering
    \includegraphics[width=0.5\textwidth]{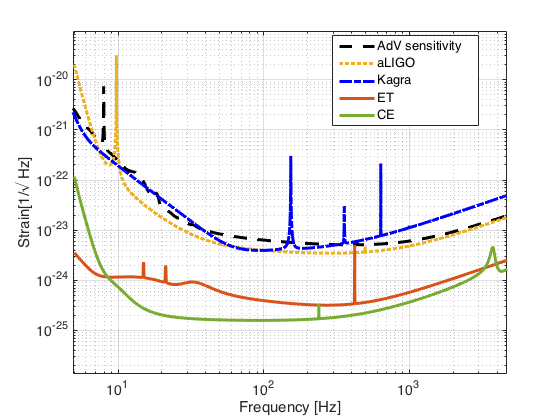}
      \caption{Sensitivity curves of the current and next generation gravitational wave laser-interferometers. We highlight that in the case of Advanced Virgo (AdV) and advanced LIGO (aLIGO) we give the design sensitivity in broadband configuration with input laser power 125W.
        \label{fig:LIsens}
}
\end{figure}

\textit{Gravitational-wave detection} --- The first detections of gravitational waves by the LIGO-Virgo network marked the beginning of gravitational-wave astronomy \cite{AbEA2016a, AbEA2016d,AbEA2017b,AbEA2017c,AbEA2017d}. While LIGO and Virgo will continue taking data in the next years, the Japanese detector KAGRA~\cite{AsEA2013} and LIGO India \cite{Unn2013} will join the network. Gradual sensitivity improvements are the result of continuous commissioning work, and more substantial technology upgrades of the current detectors are being envisaged in the mid 2020 \cite{LSC2015a, AbEA2017a}. In parallel, much more sensitive detectors, hosted in new infrastructures, are being studied: the European Einstein Telescope (ET)~\cite{ET2011}, and the US Cosmic Explorer (CE)~\cite{AbEA2017a}. The sensitivities of present and future Earth-based gravitational-wave detectors are shown Fig.~\ref{fig:LIsens}.

Current Earth-based gravitational-wave detectors are limited by seismic noise at a few Hz, since the seismic isolation is largely based on vertical pendula, which have resonance frequencies of $\sim$ 1\,Hz. However, very interesting gravitational-wave sources are expected below 1\,Hz~\cite{CoCo2004,ASFR2006}, and this is the main motivation for the gravitational-wave space missions LISA~\cite{ASEA2017} and DECIGO~\cite{SaEA2009, SaEA2017}. In parallel, efforts are ongoing to study the feasibility of sub-Hz Earth-based gravitational-wave detectors \cite{HaEA2013}. Three main concept are explored: torsion bar antennas, superconducting gravity gradiometers~\cite{MPC2002,HCW2007,PaEA2016} and laser atom interferometers~\cite{SoEA2014,HoEA2011,CaEA2016a}. 

A torsion bar is formed by two bar-shaped test masses, which are suspended orthogonally as torsion pendula, from the same suspension point. A gravity perturbation is measured through interferometric sensors monitoring the differential angular displacement of the two bars. Suspension from the same point and measurement of the differential angular displacement guarantee partial immunity to translational seismic noise. Moreover, the angular seismic noise is filtered due to the very low resonance frequency of the bar torsional mode.  

A superconducting gravity gradiometer consists of a rigid and compact frame holding SQUID transducers. Also in this case, the gravity fluctuation is measured by a differential readout of the transducer signals. The differential measurement guarantees a partial rejection of the seismic noise. For both torsion bars and superconducting gravity gradiometers, the seismic noise can be further reduced using very sensitive accelerometers and a feedback system acting on the suspension point. 

In gravity strainmeters based on atom interferometers, a laser beam interacts with spatially-separated atomic fountains. The gravity perturbation affects the phase of the laser and this affects the interaction between the laser and the clouds of atoms. The fact that the test masses here are constituted by freely falling atoms makes these detectors less sensitive to seismic perturbations than conventional laser interferometers. 

\textit{Earthquake detection} --- Earthquakes produce a redistribution of masses in a medium and at its surface leading to changes of the gravitational field. These changes propagate at the speed of light. The corresponding gravity signal can be used in addition to seismic waves currently used in earthquake early-warning systems, to assess the occurrence of an earthquake and its properties (magnitude, position, type of fault rupture) \cite{MoEA2016}. Moreover, the gravity signal can potentially bring alternative information with respect to the seismic waves. As discussed in \cite{HaEA2015}, the prompt gravity signals produced by an earthquake are potentially observable at frequencies below 1\,Hz. Since this signal after a few seconds of fault rupture is much weaker than ground acceleration in gravimeters due to ambient seismic fields, the only way to detect these so-called co-seismic gravity changes is to measure the gravity strain (or gradient), using a strainmeter similar to the detectors planned for low frequency gravitational-wave detection discussed above. Since the sensitivities needed to detect gravitational waves are much higher than the one required to detect earthquake gravity perturbations, the development of prototypes for sub-Hz gravitational-wave detectors coincides with the development of detectors potentially suitable for geophysical applications. 

\begin{figure}
  \centering
    \includegraphics[width=0.5\textwidth]{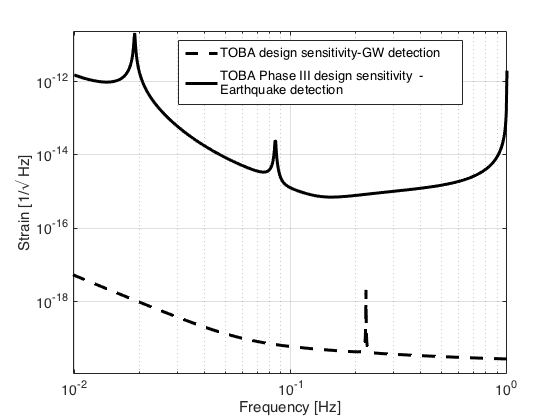}
      \caption{TOBA sensitivity curves for GW and earthquake detection. }
    \label{fig:TOBA_GWE}
\end{figure}

In the following, we will focus on the TOBA concept. In Fig.~\ref{fig:TOBA_GWE}, two TOBA detector sensitivity targets are shown. The first sensitivity (solid line) is for a medium term instrument, which is able to detect prompt gravity perturbations for earthquakes. Such an instrument would also be a prototype for a more advanced detector for gravitational-wave detection (dotted line).

\section{Sources of atmospheric gravity perturbations and their implications}
\label{sec:sources}

Before facing in detail the issues connected to the infrasound NN modeling, in this section, we review the main sources of atmospheric NN as already discussed in previous works \cite{Sau1984, Cre2008, Har2015, CaAl2009}. We will recall how atmospheric NN is produced by infrasound waves, temperature fluctuations, shockwaves and turbulent phenomena. We will also analyse the results obtained so far and their implications for gravity strainmeters.

\subsection{Infrasound waves } 
\label{sec:innoise}
NN from pressure perturbations was first treated in \cite{Sau1984} and it was further studied in \cite{Cre2008} and \cite{Har2015}. We are here addressing propagating perturbations of the mean atmospheric pressure, $p_0$, (that is sound waves), which can in turn induce density perturbations. If the perturbation of the pressure is small compared to $p_0$ and the sound wave has low frequency, the following relation holds between pressure perturbation and density perturbation~\cite{Cre2008, Har2015}: 
\begin{equation}
\gamma \frac{\delta \rho(\vec{r},t)}{\rho_0}= \frac{\delta p(\vec{r},t)}{p_0},
\end{equation}
with $\rho_0$ the mean density of the atmosphere and $\gamma \sim1.4$ the adiabatic index. Because of induced density perturbations, a local gravity potential perturbation is generated. By assuming that the infrasound wave incident on the Earth surface is perfectly reflected and by modeling the sound wave as a plane wave, the local gravity potential perturbation at a given point $\vec{r}_0$, in cylindrical coordinates, reads \cite{Har2015}:
\begin{align}
\label{eq:pin}
&\delta \phi(\vec{r}_0,t)=-\frac{G \rho_0}{\gamma p_0} \mathrm{e}^{\irm(\vec{k}_\varrho\cdot \vec{\varrho}_0-\omega t)} \delta p(\omega)  \nonumber\\
&\cdot \int_H  \mathrm{d} V \frac{ (\mathrm{e}^{\irm k_z z}-\mathrm{e}^{-\irm k_z z})\mathrm{e}^{-\irm \vec{k}_\varrho \cdot \vec{\varrho}}}{(\varrho^2 +(z-z_0)^2)^{1/2}} \nonumber\\
&=4\pi \frac{G \rho_0}{\gamma p_0} \mathrm{e}^{\irm(\vec{k}_\varrho\cdot \vec{\rho}_0-\omega t)} \nonumber\\
&\cdot \left(\mathrm{e}^{-k_\varrho \mid z_0 \mid} (2 \Theta(z_0)-1) -2 \cos(k_z z_0)\Theta(z_0)\right) \frac{\delta p(\omega)}{k^2},
\end{align}
where the integration has been performed over the whole half space $z>0$, with $z$ the vertical coordinate and $\vec\varrho$ the horizontal ones. Furthermore, $\Theta(\cdot)$ denotes the Heaviside function, and $k$ is the sound wavenumber, with $k_\varrho$ and $k_{z}$ the horizontal and vertical components, respectively. We also notice that $\vec{\rho}_0$ and $z_0$ define the horizontal coordinates and the height with respect to the Earth's surface of the considered point $\vec{r}_0$, respectively. An estimate of the infrasound NN for the current laser-interferometer detectors can be found in \cite{Har2015}. We will analyse in detail the modeling of the infrasound NN and the different aspects it implies, in section \ref{sec:inm}.

\subsection{Temperature fluctuations} 
Perturbations of the local gravity field can also be due to the temperature fluctuations in the atmosphere. Indeed, these fluctuations cause atmospheric density perturbations. Assuming small temperature changes, the ideal gas law, at constant pressure, gives the following density variation:
\begin{equation}
\delta \rho(\vec{r},t)= -\frac{\rho_0}{T_0} \delta T(\vec{r},t),
\label{eq:deltarho}
\end{equation}
where $\rho_0$ and $T_0$ are the average density and temperature of the atmosphere, respectively. It is important to stress that we address gravity perturbations due to a quasi-static temperature field whose perturbations are up-converted in frequency by advection. Given equation \ref{eq:deltarho}, the gravitational acceleration perturbation produced by this temperature field can be written as: 
\begin{equation}
\delta \vec{a}(\vec{r},t)= -\frac{G \rho_0}{T_0} \int \mathrm{d} V \frac{\delta T(\vec{r},t)}{\mid \vec{r}-\vec{r_0}\mid^3}(\vec{r}-\vec{r_0}).
\end{equation}
The difficulties to obtain an explicit expression of the temperature field requires to work with its statistical properties. If the temperature field is stationary, the spectral density associated with the gravitational acceleration perturbation reads~\cite{Har2015}: 
\begin{align}
&S(\delta \vec{a}_x; \vec{r_0},\omega)=2 \left( \frac{G \rho_0}{T_0}\right)^2 \int \mathrm{d} \tau \int \mathrm{d} V  \nonumber\\
& \int \mathrm{d} V ' \frac{x x'}{r^3(r')^3}\langle \delta T(\vec{r},t)  \delta T(\vec{r}',t+\tau)\rangle e^{\irm \omega \tau}
\end{align}
where $\vec{r}'$ and $\vec{r}$ are position vectors relative to the test-mass location. In this simplified model, the temperature field is time dependent only because the coordinate system is fixed in space (Euler coordinates). For an observer attached to an air parcel (using Lagrange coordinates), the temperature field would be static.

To perform the integration in the last equation the temperature field must be characterized. At sufficiently high frequency ($\sim$ 100 mHz), the influence of the Earth surface temperature can be neglected \cite{KuNa2006}, consequently the temperature perturbations can be considered homogeneous and isotropic. Results for the high-frequency strain noise due to the atmospheric temperature field can be found in~\cite{Cre2008} and \cite{Har2015}. In particular, these results show that for the sensitivity of the current laser-interferometer detectors (e.g. Advanced Virgo and Advanced LIGO) the NN induced by temperature fluctuations in the atmosphere should not be relevant. Nevertheless, this noise can affect the low frequency sensitivity limit of the next generation of laser-interferometer detectors (ET, CE). To obtain the strain noise at lower frequencies (below few tens of mHz), it is necessary to develop new atmospheric models which take into account the influence of the Earth surface temperature on the temperature fluctuations as well as the size of the atmosphere.

\subsection{Shockwaves} Atmospheric shockwaves can generate sudden pressure changes which can induce transient signals in the considered detectors, rather than raising their noise floor \cite{Har2015}, \cite{Cre2008}, \cite{CaAl2009}.  Atmospheric shockwaves are relevant for low frequency detectors, because they can produce significant pressure variations on time scales of order of 0.1 sec, timescale corresponding to the low frequency edge of the sensitivity curve of the majority of the current gravitational waves interferometric detectors . An example of atmospheric shockwaves are the {\em sonic booms} generated by supersonic airplanes flying over the detectors ~\cite{Cre2008, Har2015, CaAl2009}. Even though atmospheric shockwaves can effectively produce spurious signals in detectors, they are also easy to veto using environmental sensors, e.g. by placing infrasound microphones outside the buildings housing the test masses and outside the test mass vacuum chambers \cite{Cre2008}.

\subsection{Atmospheric turbulent phenomena} In what follows, the problem of atmospheric NN produced by sound radiated from a turbulent fluid flow is addressed. This subject is treated in the context of gravitational wave detectors and it is based on Lighthill's theory of the pressure-fluctuation generation in air ~\cite{Lig1952,Lig1954}. The complete calculations to obtain the gravity perturbation due to sound generated by turbulences in the atmosphere are reported in~ \cite{Har2015, CaAl2009}. The mentioned results are obtained under particular approximations that we briefly recall here: the temperature field is considered as uniform; the flow velocity components are smaller than the sound speed in the medium (i.e. small Mach number); the velocity field is stationary, isotropic and homogeneous. 

To our purpose, the important result of the calculations is that below a characteristic cutoff frequency which depends on the turbulence wavenumber and on the flow velocity components, the spectrum of the gravity perturbation is proportional to $1/\omega^2$. In addition, it is shown in~ \cite{Har2015}, that this NN is negligible above few Hz. Therefore, it does not affect current laser interferometer sensitivities, but it can be significant for the next generation of laser interferometers (ET, CE) and for ground-based sub-Hertz detectors.

\section{Infrasound NN model}
\label{sec:inm}

In this section, we describe the mathematical framework we have used to model the infrasound NN both for laser interferometers and for  TOBA. 

We consider two incoherent contributions to the atmospheric NN: the first comes from the open atmosphere above ground, and the second  from inside the buildings or the cavities housing the test masses (for a detector built on the surface or built underground, respectively). We call the first contribution \emph{exterior}, and the second contribution \emph{interior}. The underground scenario is analyzed in greater detail to investigate the suppression of atmospheric NN as a function of depth. It is worth noticing that the issue of the influence of the buildings housing the test masses has been earlier discussed in ~\cite{Cre2008}, we present here a different analysis which considers non zero pressure fluctuations inside the test mass buildings.

Starting from equation (\ref{eq:pin}), it is possible to calculate the strain noise due to atmospheric infrasound waves both for laser interferometers and for TOBA. We begin by considering the case of laser-interferometer detectors on surface ($z_0=0$). By taking into account one of the two interferometer arm test masses, its gravitational acceleration due to infrasound waves along the $x$ direction can be calculated by taking the derivative of the last expression in equation (\ref{eq:pin}) with respect to the $x$ component of the considered test mass position, i.e. $x_0$, in Cartesian coordinates. Once the gravity acceleration is obtained, it is possible to calculate the gravity strain. For an interferometer with arm length $L$, the following relation holds:
\begin{equation}
	\label{eq:LIa}
	\ddot{h}(t)=g(t)/L,
\end{equation}
with $g(t)$ the test mass gravity acceleration. Furthermore, FT$[\ddot{h}(t)]$=$-\omega^2 h(\omega)$, with FT indicating the Fourier transform. We can then calculate the one-sided power spectral density of the strain noise induced by infrasound, for a single test mass, by applying the relation $\pi S_h (\omega) \delta(\omega-\omega')=\langle h(\omega) h^* (\omega') \rangle $, where $*$ denotes the complex conjugation and $\langle...\rangle$ is the average over all the propagation directions of the infrasound wave. To obtain the total infrasound NN for a laser interferometer, we sum incoherently the spectral density of the infrasound strain noise for the four test masses. 

In order to determine the interior and exterior contributions, we need to separate the integration over the whole half space, $z>0$, of equation (\ref{eq:pin}) in two. We assume that the exterior and interior pressure fluctuations have different spectra, and that the two fields are incoherent. Furthermore, we assume that the test masses are inside vacuum chambers and there is a region around the masses itself with zero pressure  fluctuations. The required integration over the interior and exterior domains, to evaluate the gravity potential perturbation, was carried out numerically (using a MATLAB based code). 

For a laser-interferometric detector underground, in order to obtain the total atmospheric NN, we need to sum the interior contribution, computed in the same way as for a surface detector, with the exterior contribution. The latter is given by the last expression of equation (\ref{eq:pin}), with negative $z_0$, accounting for the vertical position of the test masses with respect to Earth's surface. 

Further details concerning the numerical calculations of the infrasound NN for laser interferometers will be given in the next section.

The Newtonian noise modeling for the TOBA detector configuration is slightly different from the one for laser interferometers, since Newtonian noise is highly correlated over the extent of the detector. In other words, instead of being sensitive to fluctuations of gravity acceleration coupling independently at different test masses, TOBA is sensitive to fluctuating gravity gradients.

Particularly, we start by considering the gravity potential perturbations due to atmospheric pressure fluctuations given by the last expression of equation  (\ref{eq:pin}), with $z_0=0$ for the detector on the Earth surface and $z_0<0$ for the detector placed underground.  In both cases, we then calculate the gravity gradient tensor of the gravity potential perturbations, i.e. the quantity $-\nabla \otimes \nabla \delta \phi(\vec{r} ,t)$, with $\otimes$ the dyadic product for the operator $\nabla=(\partial_x,\partial_y,\partial_z)$. This tensor is equivalent to the second time derivative of the strain tensor $\mathbf{h}(\vec{r},t)$ and it can then be used to calculate the TOBA strain noise. To this aim, the gravity-gradient tensor must be projected on a combination of unit vectors describing the TOBA strainmeter response to the gravity perturbations. Let's assume that the two bars of a TOBA detector are aligned along the $x$ and $y$ axes, whose unit vectors are $\vec{e}_1$ and $\vec{e}_2$, respectively. The projection to obtain the rotational strain, characteristic of TOBA, is then:
\begin{equation}
	h_{\times}(\vec{r},t)=\vec{e}_1 \cdot \mathbf{h}  (\vec{r},t) \cdot \vec{e}_2^\top,
\end{equation}
where $\top$ signifies the transpose of a vector. For $z_0=0$ and using equation (\ref{eq:pin}), the strain in Cartesian coordinates reads: 
\begin{equation}
\label{eq:hx0}
h^s_{\times}(\vec{r}_0,t)=-4\pi \frac{G \rho_0}{\gamma p_0} \mathrm{e}^{\irm(\vec k_\varrho\cdot \vec\varrho_0-\omega t)} \frac{k_x k_y \delta p(\omega)}{\omega^2 k^2}.
\end{equation}
Hence, by expressing the acoustic wave-vector, $\vec{k}$, in spherical coordinates, the power spectral density of the strain noise due to infrasound waves reads:
\begin{align}
\label{eq:shx0}
S_{h^s_{\times}}(\vec{r}_0,\omega)=& \left(4\pi \frac{G \rho_0}{\gamma p_0 \omega^2}\right)^2 \cdot \nonumber\\
&\langle{(\sin^2(\theta)  \sin(\phi) \cos(\phi))^2\rangle} S_{\delta p}(\omega).
\end{align}
In the previous equation, $\theta$ is the polar angle  of the considered spherical reference frame, $\phi$ is the azimuthal angle and the quantity $\langle...\rangle$ represents the average over all the acoustic wave propagation directions. The quantity $S_{\delta p}(\omega)$ is the power spectral density of the pressure fluctuations. 
Analogously, for the underground case, that is $z_0<0$, by making use of equation (\ref{eq:pin}), we obtain, in Cartesian coordinates:
\begin{equation}
\label{eq:hxm0}
h^u_{\times}(\vec{r}_0,t)=-4\pi \frac{G \rho_0}{\gamma p_0}\mathrm{e}^{k_\varrho z_0} \mathrm{e}^{\irm(\vec k_\varrho\cdot \vec\varrho_0-\omega t)} \frac{k_x k_y \delta p(\omega)}{\omega^2 k^2}.
\end{equation}
The power spectral density of the strain noise due to infrasound waves is then: 
\begin{align}
\label{eq:shxm0}
S_{h^u_{\times}}(\vec{r}_0,\omega)&=\left(4\pi \frac{G \rho_0}{\gamma p_0 \omega^2}\right)^2 \cdot \nonumber\\
 &\langle{\mathrm{e}^{2 k \sin{\theta} z_0}(\sin^2(\theta)  \sin(\phi) \cos(\phi))^2\rangle} S_{\delta p}(\omega),
\end{align}
with $\theta$, $\phi$, $\langle ... \rangle$ and $S_{\delta p}(\omega)$ having the meaning explained above. 

\section{Infrasound NN numerical simulation}
\label{sec:sim}

In this section, we analyze few features of the MATLAB based codes used for the infrasound NN calculations, describing first the case of an interferometer and then the case of a TOBA-like detector.  

We assume that the buildings housing the test masses are hemispheres and that the test masses are located at their horizontal center, inside spherical vacuum chambers, at a height $z_0 \geq 0$ above ground. To obtain the exterior contribution to the gravity potential perturbation, $\delta \phi_{\rm ext}(\vec{r}_0,t)$, we define a grid of points through which we numerically calculate the integration appearing in the $\delta \phi_{\rm ext}(\vec{r}_0,t)$ expression: 
\begin{align}
\label{eq:vout}
&\delta \phi_{\rm ext}(\vec{r}_0,t)=-\frac{G \rho_0}{\gamma p_0} \mathrm{e}^{\irm(\vec k_\varrho\cdot \vec\varrho_0-\omega t)} \delta p(\omega)  \nonumber\\
&\cdot \int_{V_{\rm ext}}  \mathrm{d} V \frac{ (\mathrm{e}^{\irm k_z z}-\mathrm{e}^{-\irm k_z z})\mathrm{e}^{\irm \vec k_\varrho\cdot \vec\varrho}}{(\varrho^2 +(z-z_0)^2)^{1/2}}.
\end{align}
The previous equation is the same as the first equality in equation \ref{eq:pin}, but with a different domain of integration being $V_{\rm ext}$ the volume of the open atmosphere without the contribution due to the space inside the hemispherical building. The entire volume above ground is taken to be a half space, i.e., we neither take into account the finite thickness of the atmosphere nor Earth's curvature and topography. In practice, to perform the integration in equation (\ref{eq:vout}), we fix the grid size to be four times the largest considered acoustic wavelength, so that the contributions beyond the grid, that we do not take into account, are negligible. We also point out that the quantity $\delta p(\omega)$, in the last equation, represents the exterior pressure fluctuations. The $x_0$-component of the gradient of $\delta \phi_{\rm ext}(\vec{r}_0,t)$ with respect to the test mass position coordinates gives the test-mass acceleration along one of the interferometer arms. Therefore, by making use of relation (\ref{eq:LIa}), the power spectral density of the strain noise due to the exterior infrasound field, for a single test mass, reads:
\begin{align}
\label{eq:svout}
&S_{\rm ext}(\vec{r}_0,\omega)=\left(\frac{G \rho_0 k}{\gamma p_0 L \omega^2}\right)^2 S^{\rm ext}_{\delta p} (\omega)\cdot  \nonumber\\
& \langle \left[\cos(\phi) \sin(\theta) \mid \mathcal{I}_{\rm ext}(z_0,\theta,\phi,\omega) \mid \right]^2 \rangle,
\end{align}
where $\mathcal{I}_{\rm ext}(z_0,\theta,\phi,\omega)$ is the integral in equation $\ref{eq:vout}$, when rewriting the components of the wave-vector $k_x$, $k_y$, $k_z$ in spherical coordinates, by using the polar angle $\theta$ and the azimuthal angle $\phi$.  As in the previous section, $\langle...\rangle$ represents the average over all the acoustic wave propagation directions and the quantity $S^{\rm ext}_{\delta p}$ is the power spectral density of the exterior pressure fluctuations.

We remark that the choice of expressing the infrasound wave-vector through spherical coordinates leads us to calculate the average in equation~\ref{eq:svout} by considering a wave-vector with spherically-uniform distribution. Hence, for the infrasound wave-vector, we consider equidistant values of the azimuthal coordinate,  $\phi_i$, in the interval $[0,2 \pi]$, and equidistant values of the function $\cos \theta_i$, between $[1,0]$, with $\theta_i$ the infrasound wave-vector polar coordinate. We evaluate the perturbation of the gravity acceleration for each couple of spherical coordinates ($\theta_i$, $\phi_i$) and then we calculate its averaged value. We also remark that the wavenumber, $k$ of the infrasound waves, for each considered frequency, $f$, is $(2 \pi f)/c_s$, with $c_s=343$ m/s, the speed of sound in air. \\

The gravity potential perturbation due to infrasound in the space inside the building, without the vacuum chamber volume, is calculated with the same method, replacing $V_{\rm ext}$ with $V_{\rm int}$, where the latter is the volume of the building without the vacuum chamber.   

In order to obtain the total infrasound NN for a single test mass, we add incoherently the interior and exterior power spectral density contributions, assuming uncorrelated pressure fluctuations in the two environments: 
\begin{equation}
\label{eq:slitot}
S_{\rm tot}(\vec{r}_0,\omega)=S_{\rm int}(\vec{r}_0,\omega)+S_{\rm ext}(\vec{r}_0,\omega).
\end{equation}
The total infrasound NN for laser interferometers is obtained by the incoherent sum of the infrasound strain noise power spectral densities for the four test masses. This is equivalent to multiply $S_{\rm tot}(\vec{r}_0,t)$ by a factor four, when assuming that the test masses have the same infrasound NN spectra. We point out that if the laser interferometers is located underground the gravity perturbation inside the underground cavity is calculated as the one inside the building on the Earth surface, with appropriate choices for $z_0$.   

For a TOBA-like detector, we have already given most of the details to estimate the infrasound NN for this strain-meter configuration, in section (\ref{sec:inm}) (see equations  \ref{eq:shx0}, \ref{eq:shxm0}). Here we underline that, in this case, we neglect the building effect since the cavity or building hosting the detector is much smaller than the length of infrasound waves, which means that the overall air mass subject to density fluctuations is simply too small to have a significant effect. As a consequence, we do not perform numerical integrations with MATLAB, because we can use the analytical solution given in equation (\ref{eq:pin}) for our calculations.  

Even if we do not integrate inside the MATLAB program, we still need to calculate the average over all the infrasound wave-vector propagation directions, which is involved in equations \ref{eq:shx0}, \ref{eq:shxm0}. The method applied in our codes to perform this calculation is analogous to the one described above, for the laser interferometers.

\section{Measurements of pressure spectra}
\label{sec:pspectra}

A key aspect of infrasound NN modeling is the measurement of pressure fluctuations, which change with time, season and place~\cite{BBB2005}. To this aim, we have made a campaign of measurements of infrasound acoustic noise at the Virgo site, in Cascina (Italy). We used Br\"{u}el \& Kj\ae r microphones, model 4193-L-004~\cite{BKmic2017}, whose response is flat up to 0.5 Hz, together with microphone conditioning amplifiers NEXUS 2690~\cite{BKnex2017} allowing a passband of -1dB from 0.1Hz to 100kHz, that is well in agreement with the microphone response.

\begin{figure}
  \centering
    \includegraphics[width=0.43\textwidth]{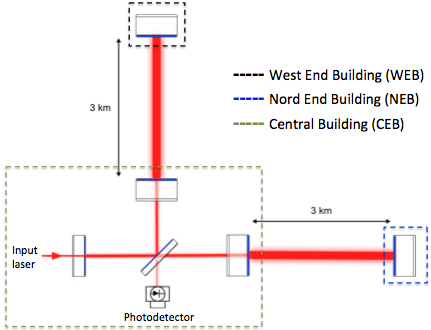}
    \caption{Simplified scheme of the AdV detector.}
 	\label{fig:AdVSScheme}
\end{figure}

\begin{figure}
  \centering
    \includegraphics[width=0.55\textwidth]{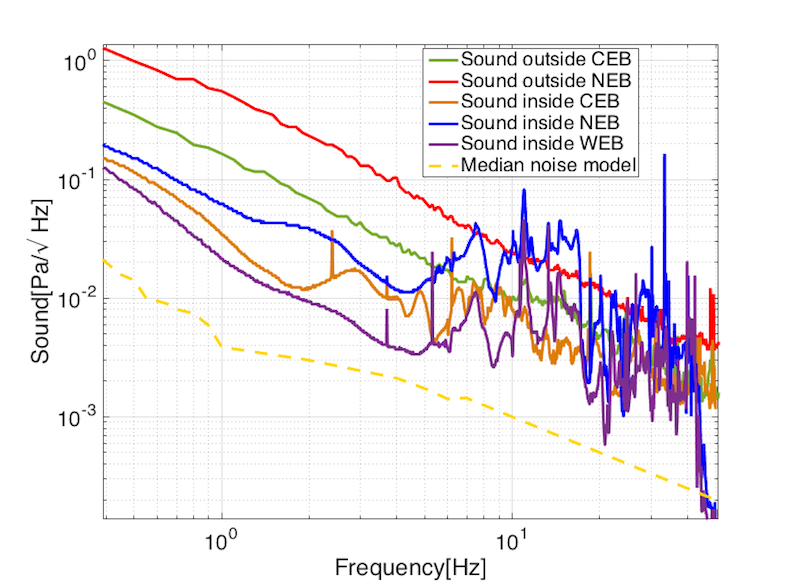}
    \caption{The solid lines correspond to the sound spectra measured at the AdV site. The dashed yellow line corresponds to the pressure fluctuation median noise model presented in~\cite{BBB2005}.}
 	\label{fig:soundAdV}
\end{figure}

We measured the sound level inside the two end-station buildings of Virgo, i.e. the North End Building (NEB) and the West End Building (WEB), and in the CEntral Building (CEB), see the AdV simplified scheme reported in Fig.~\ref{fig:AdVSScheme}. We used three months of data to estimate the acoustic noise inside each building (using data after modifications of ventilation systems end of July 2017). The corresponding 90th percentiles of the acoustic spectra for each station are reported in Fig.~\ref{fig:soundAdV}.

\begin{figure}
  \centering
    \includegraphics[width=0.55\textwidth]{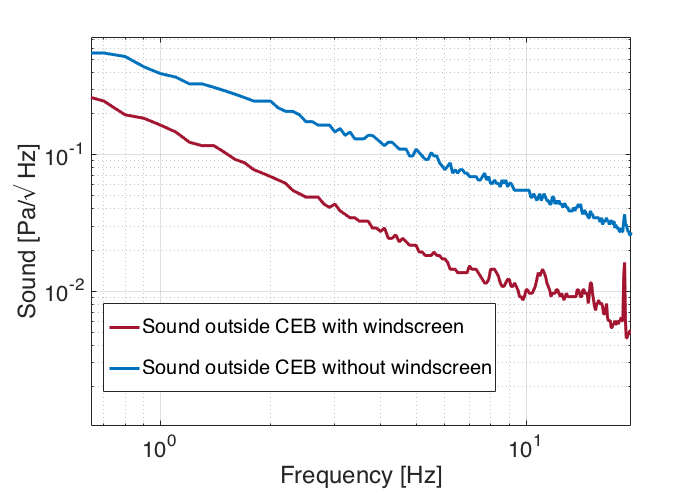}
    \caption{Sound spectra recorded outside the AdV Central Building(CEB), with and without windscreen.}
    \label{fig:soundWSCEB}
\end{figure}

\begin{figure}  
  \centering
    \includegraphics[width=0.55\textwidth]{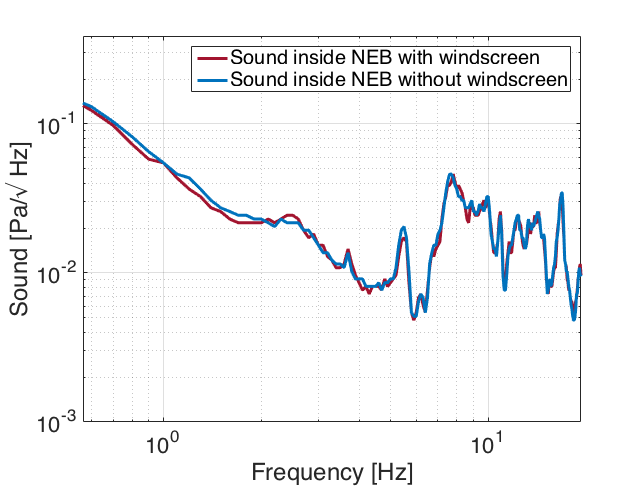}
     \caption{Sound spectra recorded inside the AdV North End Building (NEB), with and without windscreen.}
     \label{fig:soundWSNEB}
\end{figure}

As far as the measurements outside the buildings are concerned, there are no permanently installed microphones, so we recorded acoustic data sets lasting between 20 minutes and 1 hour. The sound spectra corresponding to the 90 percentile of the considered sample are shown in Fig.~\ref{fig:soundAdV}. It is important to point out that, in order to measure the sound outside the buildings, we applied a windscreen on the microphone to avoid that the direct pushing of the wind on the microphone leads to a wrong estimate of the sound noise. To highlight the windscreen effect, in Fig.~\ref{fig:soundWSCEB} and \ref{fig:soundWSNEB} we show the sound spectra recorded with and without windshield outside and inside a building, respectively. As you can see from Fig.~\ref{fig:soundWSNEB}, the windshield has a negligible effect on the recorded spectrum inside the building, where there is no wind. Fig.~\ref{fig:soundWSCEB} shows instead a significant attenuation of the acoustic noise when using the windshield, if the data are recorded outside the building. However, it might be possible that the windshield we have used is unable to completely get rid of the wind effect. Consequently, the acoustic noise outside the buildings may be lower than the one in Fig.~\ref{fig:soundAdV}, and the corresponding measurement can only be considered an upper limit. We also stress that the day during which we took the sound spectra at the CEB was less windy than the day we measured sound at NEB. This should explain, at least partially, the higher noise outside the NEB compared to the one recorded outside the CEB. Furthermore, by looking at Fig.~\ref{fig:soundAdV}, it is worth pointing out that the AdV measured spectra are mostly higher than the pressure fluctuation median noise model reported in ~\cite{BBB2005}. In particular, we observe that the AdV buildings are quite loud: the inside sound spectra at $\sim$ 10 Hz and above are close to and in some cases significantly higher than the outside spectra. This is mainly due to the ventilation systems inside the buildings.

\begin{figure}
  \centering
    \includegraphics[width=0.5\textwidth]{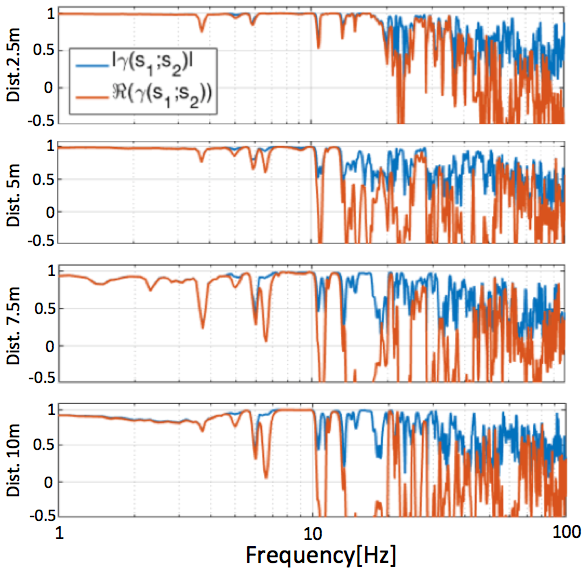}
    \caption{Sound correlation analysis inside the AdV NEB, without windshield. The microphone distances are indicated on the left of the plot. The curves shown here in each plot are the absolute value and real value of complex coherence.}
	 \label{fig:c_neb_inside}
\end{figure}

We also analyzed the correlation between sound measured by two microphones of the same type located at a fixed distance. We recorded sound for several distances of the two microphones, i.e. 2.5 m, 5 m, 7.5 m and 10 m. Each data set lasts between 20 minutes and half an hour. 
This study was carried out at the NEB, both inside the test mass building and outside. In Fig.~\ref{fig:c_neb_inside} we report the results of the analysis for the case inside the building, where sound is recorded without windshield on the microphones. We remark that, as expected, the correlation decreases both with the microphone distance and with the acoustic frequency. 

\begin{figure}
  \centering
    \includegraphics[width=0.5\textwidth]{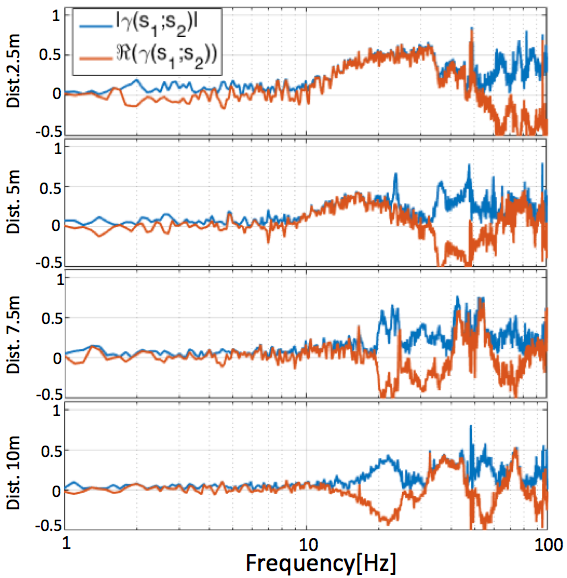}
    \caption{Sound correlation outside the AdV NEB with windshield on the two microphones. The microphone distances are indicated on the left of the plot. The curves shown here in each plot are the absolute value and real value of complex coherence.}
      \label{fig:c_out_neb}
\end{figure}

In Fig.~\ref{fig:c_out_neb} we present the correlation results for the case outside the building, when using the windshields on the microphones. In this case, we can see a non-zero correlation for frequencies between 10 Hz and 40 Hz, while it seems that at frequencies below 10 Hz there is no correlation. By considering the correlation plots outside and inside the building along with the AdV sound spectra of Fig.~\ref{fig:soundAdV}, the lack of correlation at low frequencies in Fig.~\ref{fig:c_out_neb} is understood as due to the pushing of the wind flow on the microphones. Since the day during which we took the sound spectra at CEB was less windy than the one we measured sound at NEB and since the sound spectrum outside CEB above 10\,Hz  shows features deviating from the characteristic wind-noise spectrum, we will use the outside CEB spectrum to estimate AdV infrasound NN, which provides at least a rough estimate of NN from outside the buildings. Finally, the measurement outside the CEB below $\sim 10$ Hz will be taken as an upper limit of the outside sound spectra. 

\section{Results and discussions} 
\label{sec:results}

\subsection{Ground-based interferometers}
\label{sec:gbi}
\begin{figure}
  \centering
    \includegraphics[width=0.5\textwidth]{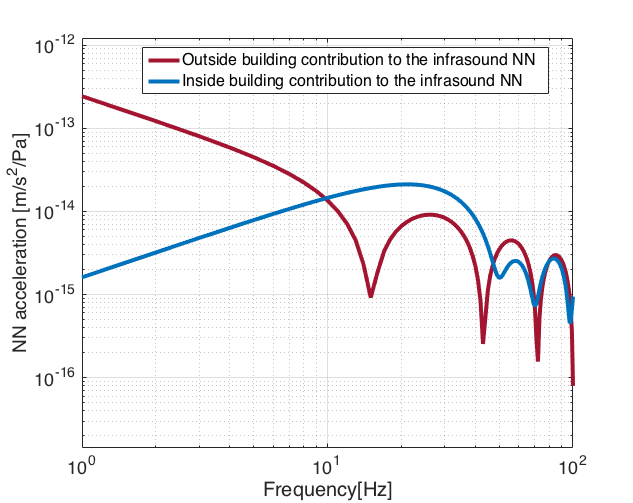}
     \caption{Contributions to the infrasound acceleration of a single test mass from the areas inside and outside the building housing the test mass.}
       \label{fig:innnac}
\end{figure}

In this section we present the noise estimations for Virgo-LIGO like detectors, taking into account the model and the pressure noise measurements described in the previous sections. We underline in particular the role of the building housing the test masses. 
In this respect, we consider a hemispherical building of 6 m radius containing a spherical vacuum chamber in its horizontal center 1.5 m above the ground. The vacuum chamber has radius of 1 m and it encloses the test mass in its center. In Fig.~\ref{fig:innnac} we show the acceleration per unit pressure of a single test mass due to the external and internal air volume. These results show that the contribution from the area inside the building becomes relevant at frequencies comparable with $c_s/D\sim 30$Hz, where $D=12$m is the building diameter and $c_S$ is the sound speed in air. At low frequencies (large wavelengths) the air inside the building does not affect significantly the test mass acceleration. The first peak of the internal contribution corresponds to a wavelength close to $D$, while the second and third peaks have wavelengths close to $D/2$ and $D/3$, respectively. We highlight that the peak patterns shown in Fig.~\ref{fig:innnac} are characteristics of the chosen building size and shape. Therefore, for different building geometrical properties the patterns would change. The total power spectral density of the infrasound strain noise for laser interferometers described in section~\ref{sec:sim}, can be obtained by using the results of Fig.~\ref{fig:innnac}.  Indeed, for a single test mass we have:
\begin{align}
\label{eq:stota}
S_{\rm tot}(\vec{r}_0,\omega)=&S_{\rm int}(\vec{r}_0,\omega)+ S_{\rm ext}(\vec{r}_0,\omega)=\nonumber\\
&\frac{1}{\left(L \omega^2 \right)^2} S^{\rm int}_{\delta p} (\omega) \langle \mid a_{\rm int}(z_0,\theta,\phi,\omega) \mid^2 \rangle+\nonumber\\
&\frac{1}{\left(L \omega^2 \right)^2} S^{\rm ext}_{\delta p} (\omega) \langle \mid a_{\rm ext}(z_0, \theta,\phi,\omega) \mid^2 \rangle,
\end{align}
where $S^{\rm int}_{\delta p} (\omega)$ and $S^{\rm ext}_{\delta p} (\omega)$ are the power spectral density of the pressure fluctuations inside and outside the test mass building, respectively. Furthermore, $\langle \mid$$a_{\rm int}(z_0, \theta,\phi, \omega)$$\mid^2 \rangle$ and $\langle \mid$$a_{\rm  ext}(z_0, \theta,\phi, \omega)$$\mid^2 \rangle$ are the square of the quantities plotted in Fig.~\ref{fig:innnac}, the former being the acceleration per unit pressure due to the internal air volume and the latter being the acceleration per unit pressure due to the external air volume.
We underline that, as in the previous sections, $\langle...\rangle$ represents the average over all the acoustic wave propagation directions given by the coordinates $(\theta, \phi)$. 

An estimate of the total infrasound NN for the AdV detector can be obtained by using the results reported in Fig.~\ref{fig:soundAdV}. To this aim, we take into account the sound spectra measured inside the three stations (CEB, NEB, WEB) and the sound spectrum measured outside the CEB. We do not use the spectrum of the acoustic noise outside the NEB, as it might still be significantly affected by the wind. From equation~\ref{eq:stota}, we can write the total power spectral density of the infrasound NN for AdV as follows: 
\begin{align}
\label{eq:stotAdV}
S^{AdV}_{\rm tot}(\omega)=&2 \times  S_{\rm tot, CEB}(\omega)+\nonumber\\
& S_{\rm tot, NEB}(\omega)+ S_{\rm tot, WEB}(\omega)=\nonumber\\
&2 \times S_{\rm int, CEB}(\omega)+ 2 \times S_{\rm ext,CEB}(\omega)+\nonumber\\
& S_{\rm int, NEB}(\omega)+ S_{\rm ext,CEB}(\omega)+\nonumber\\
&S_{\rm int, WEB}(\omega)+  S_{\rm ext,CEB}(\omega).
\end{align}

\begin{figure}
  \centering
    \includegraphics[width=0.5\textwidth]{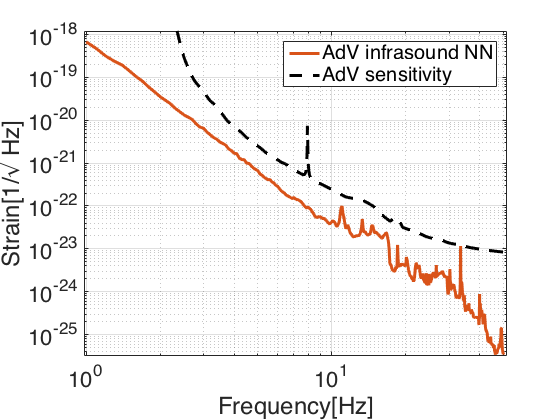}
    \caption{Estimate of the AdV infrasound NN, by using sound spectra recorded at AdV site, see Fig.~\ref{fig:soundAdV}.}
      \label{fig:innnAdV}
\end{figure}

The square root of equation~\ref{eq:stotAdV} giving the infrasound NN for Advanced Virgo is shown in Fig.~\ref{fig:innnAdV}. We observe that the infrasound NN is close to the design sensitivity between 10\,Hz and 30\,Hz, due to relatively high sound levels inside Virgo buildings. For future upgrades in this region, some care should be taken in reducing the acoustic noise inside the buildings. In addition, we stress the importance of designing less noisy HVAC systems.  


\subsection{Underground interferometers}
\label{sec:gbiu}
\begin{figure}
  \centering
    \includegraphics[width=0.5\textwidth]{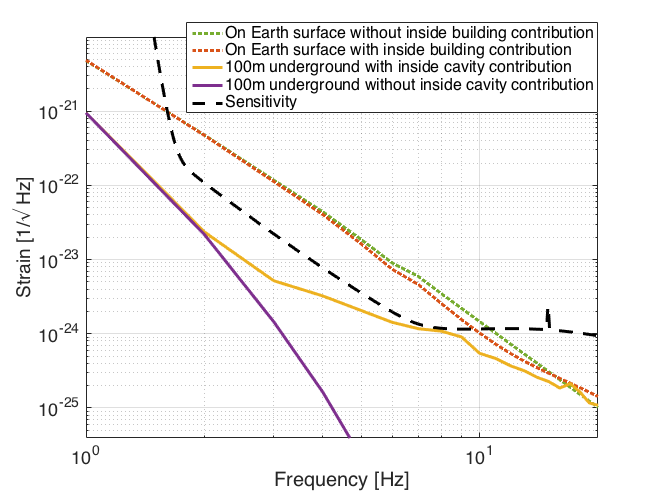}
    \caption{Infrasound NN for an ET like laser interferometer. See the dedicated section of the paper for further details.}
      \label{fig:ETINNN}
\end{figure}

In this section, we illustrate the results for an Einstein Telescope like detector and in particular we analyze the effect of underground cavities containing the detector test masses and the attenuation factor due to the detector depth beneath the earth surface. In order to highlight these aspects, we take into account four different configurations: two configurations on the surface, with and without buildings, and two configurations 100 m underground, with and without cavities. The buildings/cavities are hemispherical with 6 m radius and containing at their horizontal center a spherical vacuum chambers of 1 m radius, whose center is at height 1.5 m from the floor of the building. Each detector test mass is located in the middle of a vacuum chamber. Figure \ref{fig:ETINNN} shows the infrasound NN for the described configurations, when the pressure fluctuation median noise model given in ~\cite{BBB2005} is adopted. We do not use the spectra measured at the Virgo site since these are unusually loud. It is important to stress that we have used the same sound spectrum inside and outside the buildings. Furthermore, we have considered four test masses with identical and uncorrelated infrasound strain noise power spectral densities. 

From the results shown in Fig.~\ref{fig:ETINNN}, it appears that the infrasound NN is very close to the ET sensitivity between 4\,Hz and 10\,Hz. Going underground helps to suppress infrasound NN. However, the attenuation can be significantly spoiled by the internal contribution of the buildings. It is then important to have accurate test mass underground cavity model, when investigating the possibility of placing the detector underground. 

\subsection{Torsion bar antennas}
\label{sec:tobain}
\begin{figure}
  \centering
    \includegraphics[width=0.5\textwidth]{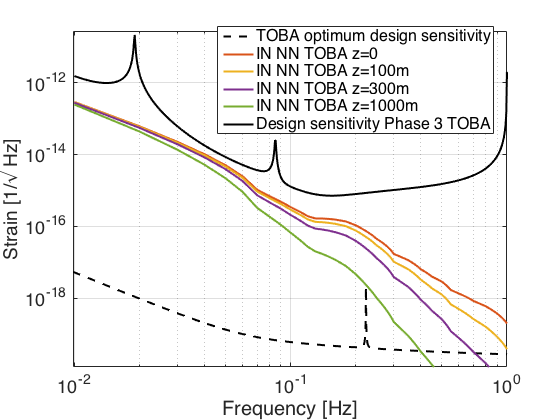}
     \caption{TOBA infrasound NN for different detector depth. The dashed sensitivity curve corresponds to the optimum TOBA configuration. The solid black curve corresponds to the next stage TOBA configuration sensitivity.}
       \label{fig:TOBAINNN}
\end{figure}

In this section, we present the results for the TOBA infrasound NN, obtained by applying the model described in sections \ref{sec:inm}, \ref{sec:sim}. The sound spectrum used for the calculations is the median-noise model reported in Fig.~6 of reference \cite{BBB2005}. 
Figure \ref{fig:TOBAINNN} shows the TOBA infrasound NN level for different detector depths together with the TOBA sensitivity curve foreseen for its ultimate GW-detector configuration (also called MANGO configuration) \cite{HaEA2013}, and the next stage TOBA sensitivity, labeled as TOBA phase three \cite{IsEA2011}. 

We point out that the capability of attenuating the infrasound NN by placing the detector underground decreases significantly at low frequencies. Indeed, the plots for different detector depths indicate that at low frequencies a larger depth beneath the earth surface is required to efficiently attenuate the infrasound NN. Furthermore, as it has been stressed, the next stage TOBA detector configuration predicts a sensitivity of $10^{-15}$m/$\sqrt{\text{Hz}}$ at 0.1Hz, for these detectors. This sensitivity is above the infrasound NN level, hence the TOBA application for geophysics purposes is not prevented by this NN. However, the sensitivity of the MANGO configuration for TOBA is considerably below the infrasound NN level. 

\begin{figure}
  \centering
    \includegraphics[width=0.5\textwidth]{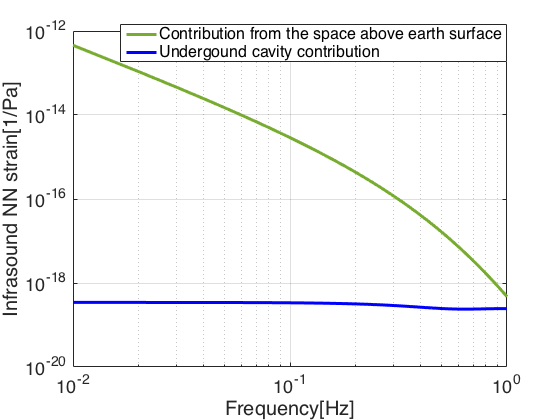}
    \caption{Contributions to the infrasound NN of a TOBA detector located 300 m beneath the earth surface. The blue line corresponds to the contribution due to the space inside the underground cavity housing the detector and the green line represents the contribution of the space above the earth surface.}
    \label{fig:TOBAINNN_un}
\end{figure}

Due to the fact that TOBA-like detectors are built to work at frequencies below 1\,Hz, the effect of the building is negligible. To better illustrate this point, in Fig.~\ref{fig:TOBAINNN_un} we report the equivalent strain per unit pressure due to the space above the earth surface and to the area inside an underground cavity, located 300 m beneath the earth surface, using the same cavity geometrical properties already considered. From Fig.~\ref{fig:TOBAINNN_un}, we can see that the contribution to the infrasound NN due to the internal air is well below the one due to the external space, even though the detector is located 300 m underground.

\section{Conclusions}
\label{sec:conclusions}
In this paper, we have addressed the problem of atmospheric gravity perturbations that affect gravitational-wave detectors as well as other gravity-gradiometric sensors. We have focused on the impact of infrasound NN on detector sensitivity in the low-frequency range (from $10^{-2}$\,Hz to 20\,Hz). We presented the mathematical framework and computational aspects of the model we have elaborated to estimate the infrasound NN level for different detector configurations. In particular we have treated the effects related to buildings and cavities housing the detector test masses, and of gravity-noise reduction by going underground. 

We have also presented new measurements of pressure noise spectra. We recall that pressure fluctuations can significantly vary with time of day, season and place \cite{BBB2005}, entailing consistent changing in the infrasound NN level. Consequently, it is important to suitably characterize the detector sites in terms of sound noise. 
 
Our model shows that infrasound NN is close to the Advanced Virgo design sensitivity, and that for future upgrades in the 10-30 Hz region a reduction of the acoustic noise in the buildings should be considered. We also find that, even if infrasound NN is strongly suppressed when going underground, the contribution  coming from the cavities housing the test masses can be significant. This contribution to infrasound NN is in fact close to the targeted sensitivity of Einstein Telescope at a few Hz. 

Experience gained with correlation measurements between microphones suggests that cancellation of infrasound NN using microphone arrays as input to Wiener filters will be very challenging. The main problem is the contribution of pressure fluctuations produced locally by wind pushing on the microphone. This deteriorates correlation between sensors and makes it unfeasible to extract the relevant information about density perturbations in the atmosphere to be able to produce a coherent estimate of associated gravity fluctuations. Cancellation of infrasound NN might therefore crucially depend on the development of alternative sensing strategies such as atmospheric LIDAR. 

For low-frequency gravitational-wave detector concepts such as TOBA that target signals below 10\,Hz (down to several tens of mHz), our model shows a level of infrasound NN below the targeted next-stage sensitivity, which is about $10^{-15}$\,m/$\sqrt{\rm Hz}$ at 0.1\,Hz. This would allow the exploitation of these instruments, for example, for geophysical applications without requiring any technology for NN cancellation. However, we predict that infrasound NN lies a few orders of magnitude above $10^{-20}$\,m/$\sqrt{\rm Hz}$ at 0.1\,Hz, a sensitivity required for gravitational-wave detection. For these low-frequency concepts, the situation does not improve significantly by going underground, which means that infrasound NN could well evolve into one of the main obstacles towards the realization of ground-based, sub-Hz gravitational-wave detectors. 

Our analysis is focused on the infrasound noise. The analysis of other sources of atmospheric NN is in progress as well as the estimation for other concepts of low frequency detectors, such as atom interferometers and superconducting gradiometers. 

\section{Acknowledgments}

We acknowledge the financial support of Agence Nationale de la Recherche through the grant E-GRAAL (ANR-14-CE03-0014-01) and the financial support of the UnivEarthS Labex program at Sorbonne Paris Cite (ANR-10-LABX-0023 and ANR-11-IDEX-0005-02).

Part of the data shown in the paper were taken using the Advanced Virgo environmental monitoring system and part were taken used specific instrumentation. 

We acknowledge the Italian Istituto Nazionale di Fisica Nucleare (INFN), the French Centre National de la Recherche Scientifique (CNRS) and the Foundation for Fundamental Research on Matter supported by the Netherlands Organisation for Scientific Research, for the construction and operation of the Virgo detector and the creation and support of the EGO consortium. 

We acknowledge colleagues from Virgo and LSC Collaborations for useful discussions and in particular Bernard Whiting.

\section{Appendix}
\begin{figure}
  \centering
    \includegraphics[width=0.55\textwidth]{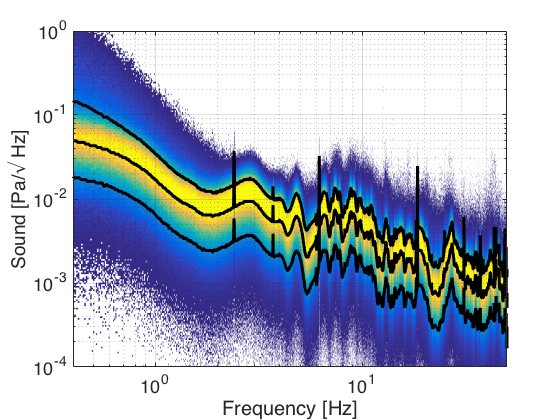}
    \caption{ Acoustic spectra inside the CEB. The black solid lines correspond to the 90th, 50th and 10th percentiles.}
 	\label{fig:soundAdV_CEBin}
\end{figure}

\begin{figure}
  \centering
    \includegraphics[width=0.55\textwidth]{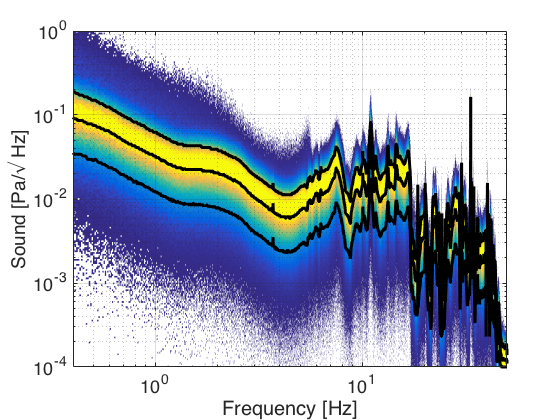}
    \caption{ Acoustic spectra inside the NEB. The black solid lines correspond to the 90th, 50th and 10th percentiles.}
 	\label{fig:soundAdV_NEBin}
\end{figure}

\begin{figure}
  \centering
    \includegraphics[width=0.55\textwidth]{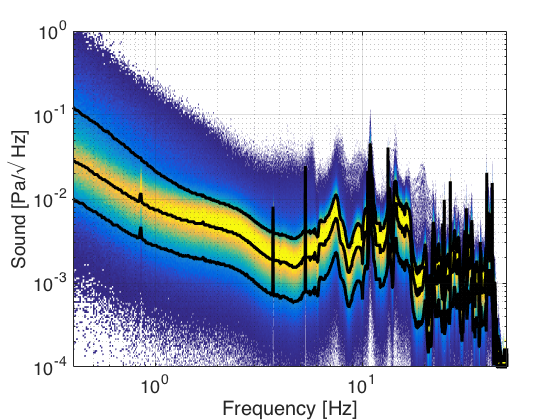}
    \caption{ Acoustic spectra inside the WEB. The black solid lines correspond to the 90th, 50th and 10th percentiles.}
 	\label{fig:soundAdV_WEBin}
\end{figure}

\begin{figure}
  \centering
    \includegraphics[width=0.55\textwidth]{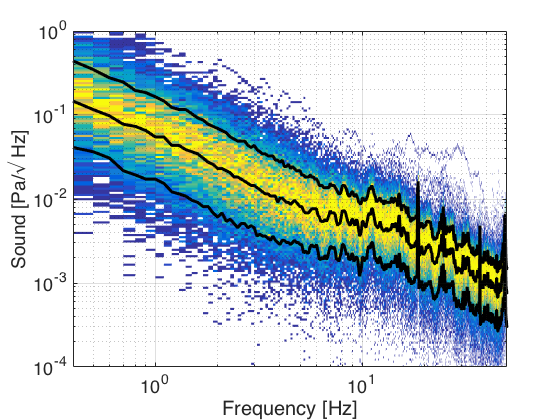}
    \caption{ Acoustic spectra outside the CEB. The black solid lines correspond to the 90th, 50th and 10th percentiles.}
 	\label{fig:soundAdVCEBout}
\end{figure}

\begin{figure}
  \centering
    \includegraphics[width=0.55\textwidth]{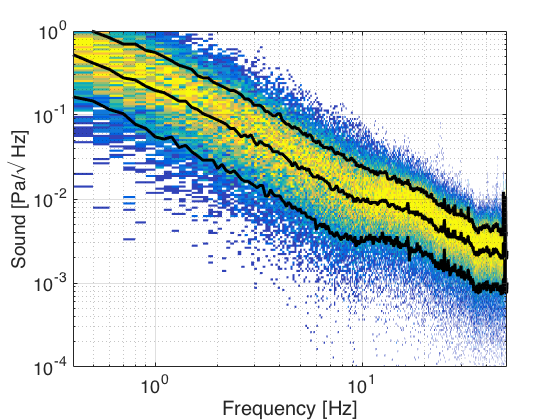}
    \caption{ Acoustic spectra outside the NEB. The black solid lines correspond to the 90th, 50th and 10th percentiles.}
 	\label{fig:soundAdVNEBout}
\end{figure}

In Fig.~\ref{fig:soundAdV_CEBin},~\ref{fig:soundAdV_NEBin} and~\ref{fig:soundAdV_WEBin}, we report spectral histograms of sound measurements inside the three stations of AdV, i.e. CEB, NEB, WEB, respectively. To obtain these plots, we used data recorded inside each building during three months, starting from the end of July 2017. The 90th, 50th and 10th percentiles of the acoustic spectra for each station are also shown. 
In Fig.~\ref{fig:soundAdVCEBout} and~\ref{fig:soundAdVNEBout}, histograms of sound measurements outside the CEB and the NEB are also presented. We stress that the observation times corresponding to these plots are shorter than the ones considered to obtain the acoustic spectra inside the buildings. In particular, we used about one hour of data to calculate the spectra outside the CEB and about half a hour of data to calculate the spectra outside the NEB.\\
We point out that the 90th percentiles shown in the five figures (Fig.~\ref{fig:soundAdV_CEBin} -~\ref{fig:soundAdVNEBout}) correspond to the curves reported in Fig.~\ref{fig:soundAdV}.

\bibliography{references}

\begin{thebibliography}{37}%
\makeatletter
\providecommand \@ifxundefined [1]{%
 \@ifx{#1\undefined}
}%
\providecommand \@ifnum [1]{%
 \ifnum #1\expandafter \@firstoftwo
 \else \expandafter \@secondoftwo
 \fi
}%
\providecommand \@ifx [1]{%
 \ifx #1\expandafter \@firstoftwo
 \else \expandafter \@secondoftwo
 \fi
}%
\providecommand \natexlab [1]{#1}%
\providecommand \enquote  [1]{``#1''}%
\providecommand \bibnamefont  [1]{#1}%
\providecommand \bibfnamefont [1]{#1}%
\providecommand \citenamefont [1]{#1}%
\providecommand \href@noop [0]{\@secondoftwo}%
\providecommand \href [0]{\begingroup \@sanitize@url \@href}%
\providecommand \@href[1]{\@@startlink{#1}\@@href}%
\providecommand \@@href[1]{\endgroup#1\@@endlink}%
\providecommand \@sanitize@url [0]{\catcode `\\12\catcode `\$12\catcode
  `\&12\catcode `\#12\catcode `\^12\catcode `\_12\catcode `\%12\relax}%
\providecommand \@@startlink[1]{}%
\providecommand \@@endlink[0]{}%
\providecommand \url  [0]{\begingroup\@sanitize@url \@url }%
\providecommand \@url [1]{\endgroup\@href {#1}{\urlprefix }}%
\providecommand \urlprefix  [0]{URL }%
\providecommand \Eprint [0]{\href }%
\providecommand \doibase [0]{http://dx.doi.org/}%
\providecommand \selectlanguage [0]{\@gobble}%
\providecommand \bibinfo  [0]{\@secondoftwo}%
\providecommand \bibfield  [0]{\@secondoftwo}%
\providecommand \translation [1]{[#1]}%
\providecommand \BibitemOpen [0]{}%
\providecommand \bibitemStop [0]{}%
\providecommand \bibitemNoStop [0]{.\EOS\space}%
\providecommand \EOS [0]{\spacefactor3000\relax}%
\providecommand \BibitemShut  [1]{\csname bibitem#1\endcsname}%
\let\auto@bib@innerbib\@empty
\bibitem [{\citenamefont {Abbott}\ \emph
  {et~al.}(2016{\natexlab{a}})\citenamefont {Abbott} \emph
  {et~al.}}]{AbEA2016a}%
  \BibitemOpen
  \bibfield  {author} {\bibinfo {author} {\bibfnamefont {B.~P.}\ \bibnamefont
  {Abbott}} \emph {et~al.} (\bibinfo {collaboration} {LIGO Scientific
  Collaboration and Virgo Collaboration}),\ }\href {\doibase
  10.1103/PhysRevLett.116.061102} {\bibfield  {journal} {\bibinfo  {journal}
  {Phys. Rev. Lett.}\ }\textbf {\bibinfo {volume} {116}},\ \bibinfo {pages}
  {061102} (\bibinfo {year} {2016}{\natexlab{a}})}\BibitemShut {NoStop}%
\bibitem [{\citenamefont {Abbott}\ \emph
  {et~al.}(2016{\natexlab{b}})\citenamefont {Abbott} \emph
  {et~al.}}]{AbEA2016d}%
  \BibitemOpen
  \bibfield  {author} {\bibinfo {author} {\bibfnamefont {B.~P.}\ \bibnamefont
  {Abbott}} \emph {et~al.} (\bibinfo {collaboration} {LIGO Scientific
  Collaboration and Virgo Collaboration}),\ }\href {\doibase
  10.1103/PhysRevLett.116.241103} {\bibfield  {journal} {\bibinfo  {journal}
  {Phys. Rev. Lett.}\ }\textbf {\bibinfo {volume} {116}},\ \bibinfo {pages}
  {241103} (\bibinfo {year} {2016}{\natexlab{b}})}\BibitemShut {NoStop}%
\bibitem [{\citenamefont {Abbott}\ \emph
  {et~al.}(2017{\natexlab{a}})\citenamefont {Abbott} \emph
  {et~al.}}]{AbEA2017b}%
  \BibitemOpen
  \bibfield  {author} {\bibinfo {author} {\bibfnamefont {B.~P.}\ \bibnamefont
  {Abbott}} \emph {et~al.} (\bibinfo {collaboration} {LIGO Scientific and Virgo
  Collaboration}),\ }\href {\doibase 10.1103/PhysRevLett.118.221101} {\bibfield
   {journal} {\bibinfo  {journal} {Phys. Rev. Lett.}\ }\textbf {\bibinfo
  {volume} {118}},\ \bibinfo {pages} {221101} (\bibinfo {year}
  {2017}{\natexlab{a}})}\BibitemShut {NoStop}%
\bibitem [{\citenamefont {Abbott}\ \emph
  {et~al.}(2017{\natexlab{b}})\citenamefont {Abbott} \emph
  {et~al.}}]{AbEA2017c}%
  \BibitemOpen
  \bibfield  {author} {\bibinfo {author} {\bibfnamefont {B.~P.}\ \bibnamefont
  {Abbott}} \emph {et~al.} (\bibinfo {collaboration} {LIGO Scientific
  Collaboration and Virgo Collaboration}),\ }\href {\doibase
  10.1103/PhysRevLett.119.141101} {\bibfield  {journal} {\bibinfo  {journal}
  {Phys. Rev. Lett.}\ }\textbf {\bibinfo {volume} {119}},\ \bibinfo {pages}
  {141101} (\bibinfo {year} {2017}{\natexlab{b}})}\BibitemShut {NoStop}%
\bibitem [{\citenamefont {Abbott}\ \emph
  {et~al.}(2017{\natexlab{c}})\citenamefont {Abbott} \emph
  {et~al.}}]{AbEA2017d}%
  \BibitemOpen
  \bibfield  {author} {\bibinfo {author} {\bibfnamefont {B.~P.}\ \bibnamefont
  {Abbott}} \emph {et~al.} (\bibinfo {collaboration} {LIGO Scientific
  Collaboration and Virgo Collaboration}),\ }\href {\doibase
  10.1103/PhysRevLett.119.161101} {\bibfield  {journal} {\bibinfo  {journal}
  {Phys. Rev. Lett.}\ }\textbf {\bibinfo {volume} {119}},\ \bibinfo {pages}
  {161101} (\bibinfo {year} {2017}{\natexlab{c}})}\BibitemShut {NoStop}%
\bibitem [{\citenamefont {Aasi}\ \emph {et~al.}(2015)\citenamefont {Aasi} \emph
  {et~al.}}]{LSC2015}%
  \BibitemOpen
  \bibfield  {author} {\bibinfo {author} {\bibfnamefont {J.}~\bibnamefont
  {Aasi}} \emph {et~al.} (\bibinfo {collaboration} {The LIGO Scientific
  Collaboration}),\ }\href {\doibase 10.1088/0264-9381/32/7/074001} {\bibfield
  {journal} {\bibinfo  {journal} {Classical and Quantum Gravity}\ }\textbf
  {\bibinfo {volume} {32}},\ \bibinfo {pages} {074001} (\bibinfo {year}
  {2015})}\BibitemShut {NoStop}%
\bibitem [{\citenamefont {Acernese}\ \emph {et~al.}(2015)\citenamefont
  {Acernese}, \citenamefont {Agathos}, \citenamefont {Agatsuma}, \citenamefont
  {Aisa}, \citenamefont {Allemandou}, \citenamefont {Allocca}, \citenamefont
  {Amarni}, \citenamefont {Astone}, \citenamefont {Balestri}, \citenamefont
  {Ballardin}, \citenamefont {Barone}, \citenamefont {Baronick}, \citenamefont
  {Barsuglia}, \citenamefont {Basti}, \citenamefont {Basti} \emph
  {et~al.}}]{AcEA2015}%
  \BibitemOpen
  \bibfield  {author} {\bibinfo {author} {\bibfnamefont {F.}~\bibnamefont
  {Acernese}}, \bibinfo {author} {\bibfnamefont {M.}~\bibnamefont {Agathos}},
  \bibinfo {author} {\bibfnamefont {K.}~\bibnamefont {Agatsuma}}, \bibinfo
  {author} {\bibfnamefont {D.}~\bibnamefont {Aisa}}, \bibinfo {author}
  {\bibfnamefont {N.}~\bibnamefont {Allemandou}}, \bibinfo {author}
  {\bibfnamefont {A.}~\bibnamefont {Allocca}}, \bibinfo {author} {\bibfnamefont
  {J.}~\bibnamefont {Amarni}}, \bibinfo {author} {\bibfnamefont
  {P.}~\bibnamefont {Astone}}, \bibinfo {author} {\bibfnamefont
  {G.}~\bibnamefont {Balestri}}, \bibinfo {author} {\bibfnamefont
  {G.}~\bibnamefont {Ballardin}}, \bibinfo {author} {\bibfnamefont
  {F.}~\bibnamefont {Barone}}, \bibinfo {author} {\bibfnamefont {J.-P.}\
  \bibnamefont {Baronick}}, \bibinfo {author} {\bibfnamefont {M.}~\bibnamefont
  {Barsuglia}}, \bibinfo {author} {\bibfnamefont {A.}~\bibnamefont {Basti}},
  \bibinfo {author} {\bibfnamefont {F.}~\bibnamefont {Basti}},  \emph
  {et~al.},\ }\href {\doibase 10.1088/0264-9381/32/2/024001} {\bibfield
  {journal} {\bibinfo  {journal} {Classical and Quantum Gravity}\ }\textbf
  {\bibinfo {volume} {32}},\ \bibinfo {pages} {024001} (\bibinfo {year}
  {2015})}\BibitemShut {NoStop}%
\bibitem [{\citenamefont {Harms}\ \emph {et~al.}(2015)\citenamefont {Harms},
  \citenamefont {Ampuero}, \citenamefont {Barsuglia}, \citenamefont
  {Chassande-Mottin}, \citenamefont {Montagner}, \citenamefont {Somala},\ and\
  \citenamefont {Whiting}}]{HaEA2015}%
  \BibitemOpen
  \bibfield  {author} {\bibinfo {author} {\bibfnamefont {J.}~\bibnamefont
  {Harms}}, \bibinfo {author} {\bibfnamefont {J.-P.}\ \bibnamefont {Ampuero}},
  \bibinfo {author} {\bibfnamefont {M.}~\bibnamefont {Barsuglia}}, \bibinfo
  {author} {\bibfnamefont {E.}~\bibnamefont {Chassande-Mottin}}, \bibinfo
  {author} {\bibfnamefont {J.-P.}\ \bibnamefont {Montagner}}, \bibinfo {author}
  {\bibfnamefont {S.~N.}\ \bibnamefont {Somala}}, \ and\ \bibinfo {author}
  {\bibfnamefont {B.~F.}\ \bibnamefont {Whiting}},\ }\href {\doibase
  10.1093/gji/ggv090} {\bibfield  {journal} {\bibinfo  {journal} {Geophysical
  Journal International}\ }\textbf {\bibinfo {volume} {201}},\ \bibinfo {pages}
  {1416} (\bibinfo {year} {2015})}\BibitemShut {NoStop}%
\bibitem [{\citenamefont {Montagner}\ \emph {et~al.}(2016)\citenamefont
  {Montagner}, \citenamefont {Juhel}, \citenamefont {Barsuglia}, \citenamefont
  {Ampuero}, \citenamefont {Chassande-Mottin}, \citenamefont {Harms},
  \citenamefont {Whiting}, \citenamefont {Bernard}, \citenamefont
  {Cl{\'e}v{\'e}d{\'e}},\ and\ \citenamefont {Lognonn{\'e}}}]{MoEA2016}%
  \BibitemOpen
  \bibfield  {author} {\bibinfo {author} {\bibfnamefont {J.-P.}\ \bibnamefont
  {Montagner}}, \bibinfo {author} {\bibfnamefont {K.}~\bibnamefont {Juhel}},
  \bibinfo {author} {\bibfnamefont {M.}~\bibnamefont {Barsuglia}}, \bibinfo
  {author} {\bibfnamefont {J.~P.}\ \bibnamefont {Ampuero}}, \bibinfo {author}
  {\bibfnamefont {E.}~\bibnamefont {Chassande-Mottin}}, \bibinfo {author}
  {\bibfnamefont {J.}~\bibnamefont {Harms}}, \bibinfo {author} {\bibfnamefont
  {B.}~\bibnamefont {Whiting}}, \bibinfo {author} {\bibfnamefont
  {P.}~\bibnamefont {Bernard}}, \bibinfo {author} {\bibfnamefont
  {E.}~\bibnamefont {Cl{\'e}v{\'e}d{\'e}}}, \ and\ \bibinfo {author}
  {\bibfnamefont {P.}~\bibnamefont {Lognonn{\'e}}},\ }\href {\doibase
  10.1038/ncomms13349} {\bibfield  {journal} {\bibinfo  {journal} {Nature
  Communications}\ }\textbf {\bibinfo {volume} {7}} (\bibinfo {year} {2016}),\
  10.1038/ncomms13349}\BibitemShut {NoStop}%
\bibitem [{\citenamefont {Saulson}(1984)}]{Sau1984}%
  \BibitemOpen
  \bibfield  {author} {\bibinfo {author} {\bibfnamefont {P.~R.}\ \bibnamefont
  {Saulson}},\ }\href {\doibase 10.1103/PhysRevD.30.732} {\bibfield  {journal}
  {\bibinfo  {journal} {Phys. Rev. D}\ }\textbf {\bibinfo {volume} {30}},\
  \bibinfo {pages} {732} (\bibinfo {year} {1984})}\BibitemShut {NoStop}%
\bibitem [{\citenamefont {Harms}(2015)}]{Har2015}%
  \BibitemOpen
  \bibfield  {author} {\bibinfo {author} {\bibfnamefont {J.}~\bibnamefont
  {Harms}},\ }\href {\doibase 10.1007/lrr-2015-3} {\bibfield  {journal}
  {\bibinfo  {journal} {Living Reviews in Relativity}\ }\textbf {\bibinfo
  {volume} {18}} (\bibinfo {year} {2015}),\ 10.1007/lrr-2015-3}\BibitemShut
  {NoStop}%
\bibitem [{\citenamefont {Creighton}(2008)}]{Cre2008}%
  \BibitemOpen
  \bibfield  {author} {\bibinfo {author} {\bibfnamefont {T.}~\bibnamefont
  {Creighton}},\ }\href {http://stacks.iop.org/0264-9381/25/i=12/a=125011}
  {\bibfield  {journal} {\bibinfo  {journal} {Classical and Quantum Gravity}\
  }\textbf {\bibinfo {volume} {25}},\ \bibinfo {pages} {125011} (\bibinfo
  {year} {2008})}\BibitemShut {NoStop}%
\bibitem [{\citenamefont {Aso}\ \emph {et~al.}(2013)\citenamefont {Aso},
  \citenamefont {Michimura}, \citenamefont {Somiya}, \citenamefont {Ando},
  \citenamefont {Miyakawa}, \citenamefont {Sekiguchi}, \citenamefont
  {Tatsumi},\ and\ \citenamefont {Yamamoto}}]{AsEA2013}%
  \BibitemOpen
  \bibfield  {author} {\bibinfo {author} {\bibfnamefont {Y.}~\bibnamefont
  {Aso}}, \bibinfo {author} {\bibfnamefont {Y.}~\bibnamefont {Michimura}},
  \bibinfo {author} {\bibfnamefont {K.}~\bibnamefont {Somiya}}, \bibinfo
  {author} {\bibfnamefont {M.}~\bibnamefont {Ando}}, \bibinfo {author}
  {\bibfnamefont {O.}~\bibnamefont {Miyakawa}}, \bibinfo {author}
  {\bibfnamefont {T.}~\bibnamefont {Sekiguchi}}, \bibinfo {author}
  {\bibfnamefont {D.}~\bibnamefont {Tatsumi}}, \ and\ \bibinfo {author}
  {\bibfnamefont {H.}~\bibnamefont {Yamamoto}} (\bibinfo {collaboration} {The
  KAGRA Collaboration}),\ }\href {\doibase 10.1103/PhysRevD.88.043007}
  {\bibfield  {journal} {\bibinfo  {journal} {Phys. Rev. D}\ }\textbf {\bibinfo
  {volume} {88}},\ \bibinfo {pages} {043007} (\bibinfo {year}
  {2013})}\BibitemShut {NoStop}%
\bibitem [{\citenamefont {Unnikrishnan}(2013)}]{Unn2013}%
  \BibitemOpen
  \bibfield  {author} {\bibinfo {author} {\bibfnamefont {C.~S.}\ \bibnamefont
  {Unnikrishnan}},\ }\href {\doibase 10.1142/S0218271813410101} {\bibfield
  {journal} {\bibinfo  {journal} {International Journal of Modern Physics D}\
  }\textbf {\bibinfo {volume} {22}},\ \bibinfo {pages} {1341010} (\bibinfo
  {year} {2013})}\BibitemShut {NoStop}%
\bibitem [{\citenamefont {{LIGO Scientific Collaboration}}(2015)}]{LSC2015a}%
  \BibitemOpen
  \bibfield  {author} {\bibinfo {author} {\bibnamefont {{LIGO Scientific
  Collaboration}}},\ }\href@noop {} {\bibfield  {journal} {\bibinfo  {journal}
  {LIGO-T1400316-v5}\ } (\bibinfo {year} {2015})}\BibitemShut {NoStop}%
\bibitem [{\citenamefont {Abbott}\ \emph
  {et~al.}(2017{\natexlab{d}})\citenamefont {Abbott} \emph
  {et~al.}}]{AbEA2017a}%
  \BibitemOpen
  \bibfield  {author} {\bibinfo {author} {\bibfnamefont {B.~P.}\ \bibnamefont
  {Abbott}} \emph {et~al.},\ }\href
  {http://stacks.iop.org/0264-9381/34/i=4/a=044001} {\bibfield  {journal}
  {\bibinfo  {journal} {Classical and Quantum Gravity}\ }\textbf {\bibinfo
  {volume} {34}},\ \bibinfo {pages} {044001} (\bibinfo {year}
  {2017}{\natexlab{d}})}\BibitemShut {NoStop}%
\bibitem [{\citenamefont {{ET Science Team}}(2011)}]{ET2011}%
  \BibitemOpen
  \bibfield  {author} {\bibinfo {author} {\bibnamefont {{ET Science Team}}},\
  }\href@noop {} {\bibfield  {journal} {\bibinfo  {journal} {{available from
  European Gravitational Observatory, document number ET-0106C-10}}\ }
  (\bibinfo {year} {2011})}\BibitemShut {NoStop}%
\bibitem [{\citenamefont {Coleman~Miller}\ and\ \citenamefont
  {Colbert}(2004)}]{CoCo2004}%
  \BibitemOpen
  \bibfield  {author} {\bibinfo {author} {\bibfnamefont {M.}~\bibnamefont
  {Coleman~Miller}}\ and\ \bibinfo {author} {\bibfnamefont {E.~J.~M.}\
  \bibnamefont {Colbert}},\ }\href {\doibase 10.1142/S0218271804004426}
  {\bibfield  {journal} {\bibinfo  {journal} {International Journal of Modern
  Physics D}\ }\textbf {\bibinfo {volume} {13}},\ \bibinfo {pages} {1}
  (\bibinfo {year} {2004})}\BibitemShut {NoStop}%
\bibitem [{\citenamefont {Amaro-Seoane}\ and\ \citenamefont
  {Freitag}(2006)}]{ASFR2006}%
  \BibitemOpen
  \bibfield  {author} {\bibinfo {author} {\bibfnamefont {P.}~\bibnamefont
  {Amaro-Seoane}}\ and\ \bibinfo {author} {\bibfnamefont {M.}~\bibnamefont
  {Freitag}},\ }\href@noop {} {\bibfield  {journal} {\bibinfo  {journal} {The
  Astrophysical Journal Letters}\ }\textbf {\bibinfo {volume} {653}},\ \bibinfo
  {pages} {L53} (\bibinfo {year} {2006})}\BibitemShut {NoStop}%
\bibitem [{\citenamefont {{Amaro-Seoane}}\ \emph {et~al.}(2017)\citenamefont
  {{Amaro-Seoane}}, \citenamefont {{Audley}}, \citenamefont {{Babak}},
  \citenamefont {{Baker}}, \citenamefont {{Barausse}}, \citenamefont
  {{Bender}}, \citenamefont {{Berti}}, \citenamefont {{Binetruy}},
  \citenamefont {{Born}}, \citenamefont {{Bortoluzzi}}, \citenamefont {{Camp}},
  \citenamefont {{Caprini}}, \citenamefont {{Cardoso}}, \citenamefont
  {{Colpi}}, \citenamefont {{Conklin}} \emph {et~al.}}]{ASEA2017}%
  \BibitemOpen
  \bibfield  {author} {\bibinfo {author} {\bibfnamefont {P.}~\bibnamefont
  {{Amaro-Seoane}}}, \bibinfo {author} {\bibfnamefont {H.}~\bibnamefont
  {{Audley}}}, \bibinfo {author} {\bibfnamefont {S.}~\bibnamefont {{Babak}}},
  \bibinfo {author} {\bibfnamefont {J.}~\bibnamefont {{Baker}}}, \bibinfo
  {author} {\bibfnamefont {E.}~\bibnamefont {{Barausse}}}, \bibinfo {author}
  {\bibfnamefont {P.}~\bibnamefont {{Bender}}}, \bibinfo {author}
  {\bibfnamefont {E.}~\bibnamefont {{Berti}}}, \bibinfo {author} {\bibfnamefont
  {P.}~\bibnamefont {{Binetruy}}}, \bibinfo {author} {\bibfnamefont
  {M.}~\bibnamefont {{Born}}}, \bibinfo {author} {\bibfnamefont
  {D.}~\bibnamefont {{Bortoluzzi}}}, \bibinfo {author} {\bibfnamefont
  {J.}~\bibnamefont {{Camp}}}, \bibinfo {author} {\bibfnamefont
  {C.}~\bibnamefont {{Caprini}}}, \bibinfo {author} {\bibfnamefont
  {V.}~\bibnamefont {{Cardoso}}}, \bibinfo {author} {\bibfnamefont
  {M.}~\bibnamefont {{Colpi}}}, \bibinfo {author} {\bibfnamefont
  {J.}~\bibnamefont {{Conklin}}},  \emph {et~al.},\ }\href@noop {} {\bibfield
  {journal} {\bibinfo  {journal} {ArXiv e-prints}\ } (\bibinfo {year}
  {2017})},\ \Eprint {http://arxiv.org/abs/1702.00786} {arXiv:1702.00786
  [astro-ph.IM]} \BibitemShut {NoStop}%
\bibitem [{\citenamefont {Sato}\ \emph {et~al.}(2009)\citenamefont {Sato},
  \citenamefont {Kawamura}, \citenamefont {Ando}, \citenamefont {Nakamura},
  \citenamefont {Tsubono}, \citenamefont {Araya}, \citenamefont {Funaki},
  \citenamefont {Ioka}, \citenamefont {Kanda}, \citenamefont {Moriwaki},
  \citenamefont {Musha}, \citenamefont {Nakazawa}, \citenamefont {Numata},
  \citenamefont {ichiro Sakai}, \citenamefont {Seto} \emph
  {et~al.}}]{SaEA2009}%
  \BibitemOpen
  \bibfield  {author} {\bibinfo {author} {\bibfnamefont {S.}~\bibnamefont
  {Sato}}, \bibinfo {author} {\bibfnamefont {S.}~\bibnamefont {Kawamura}},
  \bibinfo {author} {\bibfnamefont {M.}~\bibnamefont {Ando}}, \bibinfo {author}
  {\bibfnamefont {T.}~\bibnamefont {Nakamura}}, \bibinfo {author}
  {\bibfnamefont {K.}~\bibnamefont {Tsubono}}, \bibinfo {author} {\bibfnamefont
  {A.}~\bibnamefont {Araya}}, \bibinfo {author} {\bibfnamefont
  {I.}~\bibnamefont {Funaki}}, \bibinfo {author} {\bibfnamefont
  {K.}~\bibnamefont {Ioka}}, \bibinfo {author} {\bibfnamefont {N.}~\bibnamefont
  {Kanda}}, \bibinfo {author} {\bibfnamefont {S.}~\bibnamefont {Moriwaki}},
  \bibinfo {author} {\bibfnamefont {M.}~\bibnamefont {Musha}}, \bibinfo
  {author} {\bibfnamefont {K.}~\bibnamefont {Nakazawa}}, \bibinfo {author}
  {\bibfnamefont {K.}~\bibnamefont {Numata}}, \bibinfo {author} {\bibfnamefont
  {S.}~\bibnamefont {ichiro Sakai}}, \bibinfo {author} {\bibfnamefont
  {N.}~\bibnamefont {Seto}},  \emph {et~al.},\ }\href
  {http://stacks.iop.org/1742-6596/154/i=1/a=012040} {\bibfield  {journal}
  {\bibinfo  {journal} {Journal of Physics: Conference Series}\ }\textbf
  {\bibinfo {volume} {154}},\ \bibinfo {pages} {012040} (\bibinfo {year}
  {2009})}\BibitemShut {NoStop}%
\bibitem [{\citenamefont {Sato}\ \emph {et~al.}(2017)\citenamefont {Sato},
  \citenamefont {Kawamura}, \citenamefont {Ando}, \citenamefont {Nakamura},
  \citenamefont {Tsubono}, \citenamefont {Araya}, \citenamefont {Funaki},
  \citenamefont {Ioka}, \citenamefont {Kanda}, \citenamefont {Moriwaki},
  \citenamefont {Musha}, \citenamefont {Nakazawa}, \citenamefont {Numata},
  \citenamefont {ichiro Sakai}, \citenamefont {Seto} \emph
  {et~al.}}]{SaEA2017}%
  \BibitemOpen
  \bibfield  {author} {\bibinfo {author} {\bibfnamefont {S.}~\bibnamefont
  {Sato}}, \bibinfo {author} {\bibfnamefont {S.}~\bibnamefont {Kawamura}},
  \bibinfo {author} {\bibfnamefont {M.}~\bibnamefont {Ando}}, \bibinfo {author}
  {\bibfnamefont {T.}~\bibnamefont {Nakamura}}, \bibinfo {author}
  {\bibfnamefont {K.}~\bibnamefont {Tsubono}}, \bibinfo {author} {\bibfnamefont
  {A.}~\bibnamefont {Araya}}, \bibinfo {author} {\bibfnamefont
  {I.}~\bibnamefont {Funaki}}, \bibinfo {author} {\bibfnamefont
  {K.}~\bibnamefont {Ioka}}, \bibinfo {author} {\bibfnamefont {N.}~\bibnamefont
  {Kanda}}, \bibinfo {author} {\bibfnamefont {S.}~\bibnamefont {Moriwaki}},
  \bibinfo {author} {\bibfnamefont {M.}~\bibnamefont {Musha}}, \bibinfo
  {author} {\bibfnamefont {K.}~\bibnamefont {Nakazawa}}, \bibinfo {author}
  {\bibfnamefont {K.}~\bibnamefont {Numata}}, \bibinfo {author} {\bibfnamefont
  {S.}~\bibnamefont {ichiro Sakai}}, \bibinfo {author} {\bibfnamefont
  {N.}~\bibnamefont {Seto}},  \emph {et~al.},\ }\href
  {http://stacks.iop.org/1742-6596/840/i=1/a=012010} {\bibfield  {journal}
  {\bibinfo  {journal} {Journal of Physics: Conference Series}\ }\textbf
  {\bibinfo {volume} {840}},\ \bibinfo {pages} {012010} (\bibinfo {year}
  {2017})}\BibitemShut {NoStop}%
\bibitem [{\citenamefont {Harms}\ \emph {et~al.}(2013)\citenamefont {Harms},
  \citenamefont {Slagmolen}, \citenamefont {Adhikari}, \citenamefont {Miller},
  \citenamefont {Evans}, \citenamefont {Chen}, \citenamefont {M\"uller},\ and\
  \citenamefont {Ando}}]{HaEA2013}%
  \BibitemOpen
  \bibfield  {author} {\bibinfo {author} {\bibfnamefont {J.}~\bibnamefont
  {Harms}}, \bibinfo {author} {\bibfnamefont {B.~J.~J.}\ \bibnamefont
  {Slagmolen}}, \bibinfo {author} {\bibfnamefont {R.~X.}\ \bibnamefont
  {Adhikari}}, \bibinfo {author} {\bibfnamefont {M.~C.}\ \bibnamefont
  {Miller}}, \bibinfo {author} {\bibfnamefont {M.}~\bibnamefont {Evans}},
  \bibinfo {author} {\bibfnamefont {Y.}~\bibnamefont {Chen}}, \bibinfo {author}
  {\bibfnamefont {H.}~\bibnamefont {M\"uller}}, \ and\ \bibinfo {author}
  {\bibfnamefont {M.}~\bibnamefont {Ando}},\ }\href {\doibase
  10.1103/PhysRevD.88.122003} {\bibfield  {journal} {\bibinfo  {journal} {Phys.
  Rev. D}\ }\textbf {\bibinfo {volume} {88}},\ \bibinfo {pages} {122003}
  (\bibinfo {year} {2013})}\BibitemShut {NoStop}%
\bibitem [{\citenamefont {Moody}\ \emph {et~al.}(2002)\citenamefont {Moody},
  \citenamefont {Paik},\ and\ \citenamefont {Canavan}}]{MPC2002}%
  \BibitemOpen
  \bibfield  {author} {\bibinfo {author} {\bibfnamefont {M.}~\bibnamefont
  {Moody}}, \bibinfo {author} {\bibfnamefont {H.~J.}\ \bibnamefont {Paik}}, \
  and\ \bibinfo {author} {\bibfnamefont {E.~R.}\ \bibnamefont {Canavan}},\
  }\href {\doibase 10.1063/1.1511798} {\bibfield  {journal} {\bibinfo
  {journal} {Review of Scientific Instruments}\ }\textbf {\bibinfo {volume}
  {73}},\ \bibinfo {pages} {3957} (\bibinfo {year} {2002})}\BibitemShut
  {NoStop}%
\bibitem [{\citenamefont {Hinderer}\ \emph {et~al.}(2007)\citenamefont
  {Hinderer}, \citenamefont {Crossley},\ and\ \citenamefont
  {Warburton}}]{HCW2007}%
  \BibitemOpen
  \bibfield  {author} {\bibinfo {author} {\bibfnamefont {J.}~\bibnamefont
  {Hinderer}}, \bibinfo {author} {\bibfnamefont {D.}~\bibnamefont {Crossley}},
  \ and\ \bibinfo {author} {\bibfnamefont {R.}~\bibnamefont {Warburton}},\ }in\
  \href {\doibase http://dx.doi.org/10.1016/B978-044452748-6.00172-3} {\emph
  {\bibinfo {booktitle} {Treatise on Geophysics}}},\ \bibinfo {editor} {edited
  by\ \bibinfo {editor} {\bibfnamefont {G.}~\bibnamefont {Schubert}}}\
  (\bibinfo  {publisher} {Elsevier},\ \bibinfo {address} {Amsterdam},\ \bibinfo
  {year} {2007})\ pp.\ \bibinfo {pages} {65 -- 122}\BibitemShut {NoStop}%
\bibitem [{\citenamefont {Paik}\ \emph {et~al.}(2016)\citenamefont {Paik},
  \citenamefont {Griggs}, \citenamefont {Moody}, \citenamefont {Venkateswara},
  \citenamefont {Lee}, \citenamefont {Nielsen}, \citenamefont {Majorana},\ and\
  \citenamefont {Harms}}]{PaEA2016}%
  \BibitemOpen
  \bibfield  {author} {\bibinfo {author} {\bibfnamefont {H.~J.}\ \bibnamefont
  {Paik}}, \bibinfo {author} {\bibfnamefont {C.~E.}\ \bibnamefont {Griggs}},
  \bibinfo {author} {\bibfnamefont {M.~V.}\ \bibnamefont {Moody}}, \bibinfo
  {author} {\bibfnamefont {K.}~\bibnamefont {Venkateswara}}, \bibinfo {author}
  {\bibfnamefont {H.~M.}\ \bibnamefont {Lee}}, \bibinfo {author} {\bibfnamefont
  {A.~B.}\ \bibnamefont {Nielsen}}, \bibinfo {author} {\bibfnamefont
  {E.}~\bibnamefont {Majorana}}, \ and\ \bibinfo {author} {\bibfnamefont
  {J.}~\bibnamefont {Harms}},\ }\href
  {http://stacks.iop.org/0264-9381/33/i=7/a=075003} {\bibfield  {journal}
  {\bibinfo  {journal} {Classical and Quantum Gravity}\ }\textbf {\bibinfo
  {volume} {33}},\ \bibinfo {pages} {075003} (\bibinfo {year}
  {2016})}\BibitemShut {NoStop}%
\bibitem [{\citenamefont {Sorrentino}\ \emph {et~al.}(2014)\citenamefont
  {Sorrentino}, \citenamefont {Bodart}, \citenamefont {Cacciapuoti},
  \citenamefont {Lien}, \citenamefont {Prevedelli}, \citenamefont {Rosi},
  \citenamefont {Salvi},\ and\ \citenamefont {Tino}}]{SoEA2014}%
  \BibitemOpen
  \bibfield  {author} {\bibinfo {author} {\bibfnamefont {F.}~\bibnamefont
  {Sorrentino}}, \bibinfo {author} {\bibfnamefont {Q.}~\bibnamefont {Bodart}},
  \bibinfo {author} {\bibfnamefont {L.}~\bibnamefont {Cacciapuoti}}, \bibinfo
  {author} {\bibfnamefont {Y.-H.}\ \bibnamefont {Lien}}, \bibinfo {author}
  {\bibfnamefont {M.}~\bibnamefont {Prevedelli}}, \bibinfo {author}
  {\bibfnamefont {G.}~\bibnamefont {Rosi}}, \bibinfo {author} {\bibfnamefont
  {L.}~\bibnamefont {Salvi}}, \ and\ \bibinfo {author} {\bibfnamefont {G.~M.}\
  \bibnamefont {Tino}},\ }\href {\doibase 10.1103/PhysRevA.89.023607}
  {\bibfield  {journal} {\bibinfo  {journal} {Phys. Rev. A}\ }\textbf {\bibinfo
  {volume} {89}},\ \bibinfo {pages} {023607} (\bibinfo {year}
  {2014})}\BibitemShut {NoStop}%
\bibitem [{\citenamefont {Hohensee}\ \emph {et~al.}(2011)\citenamefont
  {Hohensee}, \citenamefont {Lan}, \citenamefont {Houtz}, \citenamefont {Chan},
  \citenamefont {Estey}, \citenamefont {Kim}, \citenamefont {Kuan},\ and\
  \citenamefont {M{\"u}ller}}]{HoEA2011}%
  \BibitemOpen
  \bibfield  {author} {\bibinfo {author} {\bibfnamefont {M.}~\bibnamefont
  {Hohensee}}, \bibinfo {author} {\bibfnamefont {S.}~\bibnamefont {Lan}},
  \bibinfo {author} {\bibfnamefont {R.}~\bibnamefont {Houtz}}, \bibinfo
  {author} {\bibfnamefont {C.}~\bibnamefont {Chan}}, \bibinfo {author}
  {\bibfnamefont {B.}~\bibnamefont {Estey}}, \bibinfo {author} {\bibfnamefont
  {G.}~\bibnamefont {Kim}}, \bibinfo {author} {\bibfnamefont {P.}~\bibnamefont
  {Kuan}}, \ and\ \bibinfo {author} {\bibfnamefont {H.}~\bibnamefont
  {M{\"u}ller}},\ }\href@noop {} {\bibfield  {journal} {\bibinfo  {journal}
  {General Relativity and Gravitation}\ }\textbf {\bibinfo {volume} {43}},\
  \bibinfo {pages} {1905} (\bibinfo {year} {2011})}\BibitemShut {NoStop}%
\bibitem [{\citenamefont {Canuel}\ \emph {et~al.}(2016)\citenamefont {Canuel},
  \citenamefont {Pelisson}, \citenamefont {Amand}, \citenamefont {Bertoldi},
  \citenamefont {Cormier}, \citenamefont {Fang}, \citenamefont {Gaffet},
  \citenamefont {Geiger}, \citenamefont {Harms}, \citenamefont {Holleville},
  \citenamefont {Landragin}, \citenamefont {LefÃ¨vre}, \citenamefont
  {Lhermite}, \citenamefont {Mielec}, \citenamefont {Prevedelli} \emph
  {et~al.}}]{CaEA2016a}%
  \BibitemOpen
  \bibfield  {author} {\bibinfo {author} {\bibfnamefont {B.}~\bibnamefont
  {Canuel}}, \bibinfo {author} {\bibfnamefont {S.}~\bibnamefont {Pelisson}},
  \bibinfo {author} {\bibfnamefont {L.}~\bibnamefont {Amand}}, \bibinfo
  {author} {\bibfnamefont {A.}~\bibnamefont {Bertoldi}}, \bibinfo {author}
  {\bibfnamefont {E.}~\bibnamefont {Cormier}}, \bibinfo {author} {\bibfnamefont
  {B.}~\bibnamefont {Fang}}, \bibinfo {author} {\bibfnamefont {S.}~\bibnamefont
  {Gaffet}}, \bibinfo {author} {\bibfnamefont {R.}~\bibnamefont {Geiger}},
  \bibinfo {author} {\bibfnamefont {J.}~\bibnamefont {Harms}}, \bibinfo
  {author} {\bibfnamefont {D.}~\bibnamefont {Holleville}}, \bibinfo {author}
  {\bibfnamefont {A.}~\bibnamefont {Landragin}}, \bibinfo {author}
  {\bibfnamefont {G.}~\bibnamefont {LefÃ¨vre}}, \bibinfo {author}
  {\bibfnamefont {J.}~\bibnamefont {Lhermite}}, \bibinfo {author}
  {\bibfnamefont {N.}~\bibnamefont {Mielec}}, \bibinfo {author} {\bibfnamefont
  {M.}~\bibnamefont {Prevedelli}},  \emph {et~al.}\ }(\bibinfo {year} {2016})\
  pp.\ \bibinfo {pages} {990008--990008--12}\BibitemShut {NoStop}%
\bibitem [{\citenamefont {Cafaro}\ and\ \citenamefont {Ali}(2009)}]{CaAl2009}%
  \BibitemOpen
  \bibfield  {author} {\bibinfo {author} {\bibfnamefont {C.}~\bibnamefont
  {Cafaro}}\ and\ \bibinfo {author} {\bibfnamefont {S.}~\bibnamefont {Ali}},\
  }\href@noop {} {\  (\bibinfo {year} {2009})},\ \Eprint
  {http://arxiv.org/abs/0906.4844} {arXiv:0906.4844 [gr-qc]} \BibitemShut
  {NoStop}%
\bibitem [{\citenamefont {Kukharets}\ and\ \citenamefont
  {Nalbandyan}(2006)}]{KuNa2006}%
  \BibitemOpen
  \bibfield  {author} {\bibinfo {author} {\bibfnamefont {V.}~\bibnamefont
  {Kukharets}}\ and\ \bibinfo {author} {\bibfnamefont {H.}~\bibnamefont
  {Nalbandyan}},\ }\href {\doibase 10.1134/S0001433806040050} {\bibfield
  {journal} {\bibinfo  {journal} {Izvestiya, Atmospheric and Oceanic Physics}\
  }\textbf {\bibinfo {volume} {42}},\ \bibinfo {pages} {456} (\bibinfo {year}
  {2006})}\BibitemShut {NoStop}%
\bibitem [{\citenamefont {Lighthill}(1952)}]{Lig1952}%
  \BibitemOpen
  \bibfield  {author} {\bibinfo {author} {\bibfnamefont {M.~J.}\ \bibnamefont
  {Lighthill}},\ }\href {\doibase 10.1098/rspa.1952.0060} {\bibfield  {journal}
  {\bibinfo  {journal} {Proceedings of the Royal Society of London A:
  Mathematical, Physical and Engineering Sciences}\ }\textbf {\bibinfo {volume}
  {211}},\ \bibinfo {pages} {564} (\bibinfo {year} {1952})}\BibitemShut
  {NoStop}%
\bibitem [{\citenamefont {Lighthill}(1954)}]{Lig1954}%
  \BibitemOpen
  \bibfield  {author} {\bibinfo {author} {\bibfnamefont {M.~J.}\ \bibnamefont
  {Lighthill}},\ }\href {\doibase 10.1098/rspa.1954.0049} {\bibfield  {journal}
  {\bibinfo  {journal} {Proceedings of the Royal Society of London A:
  Mathematical, Physical and Engineering Sciences}\ }\textbf {\bibinfo {volume}
  {222}},\ \bibinfo {pages} {1} (\bibinfo {year} {1954})}\BibitemShut {NoStop}%
\bibitem [{\citenamefont {Bowman}\ \emph {et~al.}(2005)\citenamefont {Bowman},
  \citenamefont {Baker},\ and\ \citenamefont {Bahavar}}]{BBB2005}%
  \BibitemOpen
  \bibfield  {author} {\bibinfo {author} {\bibfnamefont {J.~R.}\ \bibnamefont
  {Bowman}}, \bibinfo {author} {\bibfnamefont {G.~E.}\ \bibnamefont {Baker}}, \
  and\ \bibinfo {author} {\bibfnamefont {M.}~\bibnamefont {Bahavar}},\ }\href
  {\doibase 10.1029/2005GL022486} {\bibfield  {journal} {\bibinfo  {journal}
  {Geophysical Research Letters}\ }\textbf {\bibinfo {volume} {32}},\ \bibinfo
  {pages} {n/a} (\bibinfo {year} {2005})},\ \bibinfo {note}
  {l09803}\BibitemShut {NoStop}%
\bibitem [{BKm(2017)}]{BKmic2017}%
  \BibitemOpen
  \href {https://www.bksv.com/en/products/transducers/acoustic/
  microphones/microphone-preamplifier-combinations/4193-L-004} {\bibfield
  {journal} {\bibinfo  {journal} {Br{\"u}el \& Kj{\ae}r microphone model
  4193-L-004}\ } (\bibinfo {year} {2017})},\ \Eprint
  {http://arxiv.org/abs/https://www.bksv.com/en/products/transducers/acoustic/
  microphones/microphone-preamplifier-combinations/4193-L-004}
  {https://www.bksv.com/en/products/transducers/acoustic/
  microphones/microphone-preamplifier-combinations/4193-L-004} \BibitemShut
  {NoStop}%
\bibitem [{BKn(2017)}]{BKnex2017}%
  \BibitemOpen
  \href {https://www.bksv.com/media/doc/bp1702.pdf} {\bibfield  {journal}
  {\bibinfo  {journal} {Br{\"u}el \& Kj{\ae}r amplifier NEXUS 2690}\ }
  (\bibinfo {year} {2017})},\ \Eprint
  {http://arxiv.org/abs/https://www.bksv.com/media/doc/bp1702.pdf}
  {https://www.bksv.com/media/doc/bp1702.pdf} \BibitemShut {NoStop}%
\bibitem [{\citenamefont {Ishidoshiro}\ \emph {et~al.}(2011)\citenamefont
  {Ishidoshiro}, \citenamefont {Ando}, \citenamefont {Takamori}, \citenamefont
  {Takahashi}, \citenamefont {Okada}, \citenamefont {Matsumoto}, \citenamefont
  {Kokuyama}, \citenamefont {Kanda}, \citenamefont {Aso},\ and\ \citenamefont
  {Tsubono}}]{IsEA2011}%
  \BibitemOpen
  \bibfield  {author} {\bibinfo {author} {\bibfnamefont {K.}~\bibnamefont
  {Ishidoshiro}}, \bibinfo {author} {\bibfnamefont {M.}~\bibnamefont {Ando}},
  \bibinfo {author} {\bibfnamefont {A.}~\bibnamefont {Takamori}}, \bibinfo
  {author} {\bibfnamefont {H.}~\bibnamefont {Takahashi}}, \bibinfo {author}
  {\bibfnamefont {K.}~\bibnamefont {Okada}}, \bibinfo {author} {\bibfnamefont
  {N.}~\bibnamefont {Matsumoto}}, \bibinfo {author} {\bibfnamefont
  {W.}~\bibnamefont {Kokuyama}}, \bibinfo {author} {\bibfnamefont
  {N.}~\bibnamefont {Kanda}}, \bibinfo {author} {\bibfnamefont
  {Y.}~\bibnamefont {Aso}}, \ and\ \bibinfo {author} {\bibfnamefont
  {K.}~\bibnamefont {Tsubono}},\ }\href {\doibase
  10.1103/PhysRevLett.106.161101} {\bibfield  {journal} {\bibinfo  {journal}
  {Phys. Rev. Lett.}\ }\textbf {\bibinfo {volume} {106}},\ \bibinfo {pages}
  {161101} (\bibinfo {year} {2011})}\BibitemShut {NoStop}%
\end{thebibliography}%

\end{document}